\documentclass{ptephy_v1}
\usepackage{boites}
\usepackage{amsmath}
\usepackage{graphics}

\begin{document}
%%%%%%%%%%%%%%%%%%%%%%%%%%%%%%%%%%%%%%%%%%%%%%%%

\title{Proton Structure and PHENIX Experiment}

\author{\name{\fname{Jian-Wei} \surname{Qiu}}{1,2}}

\address{\affil{1}{Department of Physics, Brookhaven National Laboratory, Upton, NY 11973, USA}
\affil{2}{C.N.Yang Institute for Theoretical Physics and Department of Physics and Astronomy, 
Stony Brook University, Stony Brook, NY 11794, USA}
\email{jqiu@bnl.gov}}

\begin{abstract}%
We briefly summarize the important and critical roles that PHENIX Experiment has played 
in determining the proton's internal structure in terms of quarks and gluons, and their dynamics.  
Some pioneering measurements by PHENIX Experiment on the motion and polarization of 
quarks and gluons, as well as their correlations inside a fast moving proton are presented.
Some future opportunities and potentials of PHENIX Experiment are also discussed.
\end{abstract}

\subjectindex{QCD, Proton Structure, PHENIX}

\maketitle

%\tableofcontents

%%%%%%%%%%%%%%%%%%%%%%%%%%%%%%%%%%%%%%%%%%%%%%%%
\section{Introduction}
\label{sec:intro}
%%%%%%%%%%%%%%%%%%%%%%%%%%%%%%%%%%%%%%%%%%%%%%%%

Proton, along with neutron, known as the nucleon, is the fundamental building block 
of all atomic nuclei that make up all visible matter in the universe, 
including the stars, planets, and us.  
Proton is also known as a strongly interacting, relativistic bound state of 
quarks and gluons (referred as partons) in Quantum Chromo-Dynamics (QCD) 
\cite{Brambilla:2014jmp}. 
In order to explain the origin, the evolution, and the structure of the visible world, 
it is necessary to explore and to understand the proton's internal structure and its formation. 

The advances of accelerator technology in the last century made it possible 
to discover the quarks, the fundamental constituents of the proton,
in the nineteen sixties at Stanford Linear Accelerator Center (SLAC) 
\cite{Bloom:1969kc,Breidenbach:1969kd},
which led to the discovery of QCD.  QCD is the theory of strong interacting color charges 
that is responsible for binding colored quarks into the color neutral proton 
by the exchange of gluons \cite{Fritzsch:1973pi}.
In contrast to the quantum electromagnetism, where the force carrying photons 
are electrically neutral, the force carrying gluons in QCD carry the color charge 
causing them to interact among themselves, which is the defining property of QCD
responsible for the complex and extremely rich, while mysterious structure of the proton,
and the countless intriguing phenomena of strong interaction physics. 

Proton structure is not static, but, dynamic, full of features 
beyond our current knowledge and imagination.
Quarks and gluons interact strongly when they are far apart -- the confinement of color,
but, weakly when they are closer -- the Asymptotic Freedom; 
they appear and disappear in numbers almost continuously; and 
they are confined to form the proton while moving relativistically \cite{Brambilla:2014jmp}.
Understanding completely the proton structure and its properties 
(such as its mass, spin, size, and etc.), 
the emergent phenomena of QCD dynamics,  
is still beyond the capability of the best minds in the world today.  
It is the great intellectual challenge to explore and to understand the proton structure 
and its properties without being able to see the quarks and gluons in isolation.

Relativistic heavy ion collider (RHIC) at Brookhaven National Laboratory (BNL) 
was the first collider facility in the world to be able to perform experiments 
with the collisions of two relativistic heavy ion beams, as well as 
two polarized proton beams \cite{RHIChomepage}.  From collisions of two gold ion beams, 
RHIC discovered, later confirmed by the experiments at the Large Hadron Collider (LHC)
at CERN, the strongly interacting quark-gluon plasma (QGP),
the matter is expected to be only existed in the universe a few microseconds 
after its birth \cite{Gyulassy:2004zy}.  
From collisions of two very energetic polarized proton beams, 
RHIC, as the first and only polarized hadron collider in the world, discovered that 
from the momentum region accessible by RHIC experiments, 
gluons inside a longitudinally polarized fast moving proton are polarized and  
have a net positive polarization along the direction of the proton, equal to  
about 20\% of the proton's helicity \cite{Adamczyk:2014ozi,Adare:2014hsq}.  
With one proton beam polarized longitudinally, taking the advantage of the parity 
violation nature of the weak interaction, RHIC has performed the best measurements of 
the net sea quark polarization and its flavor dependence inside a polarized proton
\cite{Adamczyk:2014xyw,Adare:2010xa}.
With one proton beam polarized transversely, RHIC has provided and will continue providing 
tremendous new opportunities to probe and to study the quantum correlations 
between the spin direction of the proton and the preference in direction of the confined 
motion of quarks and gluons within the proton \cite{Abelev:2008af,Adare:2013ekj}.  
Such quantum correlations, without any doubt, are the most sensitive to 
the QCD dynamics and the formation of its bound states. 

The experimental measurements and discoveries by the PHENIX collaboration, 
along with those by the STAR collaboration and early BRAHMS and PHOBOS collaborations, 
have defined the RHIC science program and will continue to do so.  
In this short article, we briefly summarize the important and critical roles that 
PHENIX Experiment has played in determining the proton's internal structure,
especially emphasizing the recent achievements.  
Some future opportunities and potentials of PHENIX Experiment for exploring 
the proton structure are also discussed.  

%%%%%%%%%%%%%%%%%%%%%%%%%%%%%%%%%%%%%%%%%%%%%%%%
\section{The helicity structure of the proton}
\label{sec:pdfs}
%%%%%%%%%%%%%%%%%%%%%%%%%%%%%%%%%%%%%%%%%%%%%%%%

Parton distribution and correlation functions describe the fascinating relation
between a fast moving proton in high energy scattering and the quarks and gluons within it.
They carry rich information on proton's mysterious partonic structure that
cannot be calculated by QCD perturbation theory.
Parton distribution functions (PDFs) are the simplest of all correlation functions,
$f_{i/p}(x,\mu^2)$, defined as the probability distributions to find a quark,
an antiquark, or a gluon ($i=q,\bar{q},g$) in the proton
carrying its momentum fraction between $x$ and $x+dx$,
probed at the factorization scale $\mu$.
PDFs also play an essential role to connect the measured cross sections of colliding proton(s) 
to the short-distance scattering between quarks and gluons.
Without them, we would not be able to understand the hard probes,
cross sections with large momentum transfers, in high energy hadronic collisions,
as well as the discovery of Higgs particles in proton-proton collisions at the LHC.
PDFs are nonperturbative, but, universal, and have been traditionally
extracted from QCD global analysis of all existing high energy scattering data
in the framework of QCD factorization\cite{Gao:2013xoa,Martin:2009iq,Ball:2011mu,Alekhin:2013nda}.
The excellent agreement between the theory and data on the scaling-violation behavior, 
or the factorization scale $\mu$-dependence, of the PDFs 
has provided one of the most stringent tests for QCD as the theory of strong interaction.

Proton has spin $1/2$ in the unit of fundamental constant $\hbar$.
That is, proton has a quantized finite angular momentum when it is at the rest.  
Proton is a composite particle made of quarks and gluons.  
Its spin, like its mass and other properties, is an emergent phenomenon of QCD dynamics.  
We do not fully understand QCD if we do not understand how quarks and gluons, 
and their dynamics make up the proton's spin.

The proton's spin 1/2 was once attributed to the superposition of quark spin 
of three valence quarks according to the Quark Model.
The proton spin puzzle (also known as the spin ``crisis'') -- 
a very little of proton's spin is carried by its quarks, 
discovered by European Muon Collaboration (EMC) about
30 years ago~\cite{Ashman:1987hv,Ashman:1989ig}, 
has led to tremendous theoretical and experimental activities to 
explore the proton's spin structure and to search for an answer to one 
of the most fundamental questions in QCD -- how do the spin, the confined motion
and the dynamics of quarks and gluons inside the proton make up its spin 1/2?

%======================================================================
\subsection{The RHIC spin program}
\label{subsec:rhicspin}
%======================================================================
 
The RHIC spin program, with the only polarized proton-proton collider in the world, 
was designed to probe the internal quark/gluon structure of a polarized proton, and 
to search for the origin of and the answer to the long-standing spin puzzle.  
The RHIC spin program has the following three goals \cite{Aschenauer:2013woa}:
\begin{itemize}
\item
precision measurement of the polarized gluon distribution $g(x,\mu^2)$ 
over a large range of momentum fraction $x$ to constrain 
the gluon spin contribution to the proton's spin,
\item
measurements of the polarized quark and anti-quark flavor structure in the proton, and 
\item
studies of the novel transverse-spin phenomena in QCD.
\end{itemize}
Having the collisions of two high energy polarized proton beams with the polarization in 
either longitudinal or transverse direction, the RHIC spin program has a 
special advantage in probing the content of polarized gluons inside a polarized proton 
and the novel transverse spin phenomena in QCD.  
With its reach in both energy and polarization, the RHIC spin program 
has an unmatched capability to probe the flavor dependence of polarized
sea of the proton by studying the parity violating single-spin asymmetry of $W^\pm$-boson production.  
The RHIC spin program is playing a unique and very critical role in our effort to solve the 
proton spin puzzle.

Proton is a composite particle of quarks and gluons, and its spin could receive contribution
from the spin of quarks and gluons, as well as the orbital angular momenta of quarks and gluons
due to their confined motion.  Proton is also a highly dynamic quantum system of quarks and gluons
full of fluctuations at all distance scales. 
Numbers of quarks and gluons vary at all time, given by the probability distributions 
to find them, such as the PDFs.  
Similarly, the quark and gluon spin contributions to a polarized proton are given by
their helicity distributions, 
\begin{equation}
\Delta f_{i/p}(x,\mu^2) 
\equiv  f_{i/p}^+(x,\mu^2) - f_{i/p}^-(x,\mu^2) \, ,
\label{eq:pdf-p}
\end{equation}
defined as the difference of the probability distribution to find a parton of flavor $i$ 
with the same helicity as the proton, $f^{+}_{i,p}$, and that of opposite helicity, $f^{-}_{i,p}$.  
The spin 1/2 of a longitudinally polarized proton could be decomposed 
according to the following fundamental sum 
rule~\cite{Jaffe:1989jz,Ji:1996ek,Wakamatsu:2014zza,Ji:2012sj},
\begin{equation} 
\label{eq:ssr} 
S_p=\frac{1}{2}=\frac{1}{2}\, \Delta\Sigma(\mu^2) + \Delta G(\mu^2)  
+ \left( L_q(\mu^2)  + L_g(\mu^2)  \right) ,
\end{equation} 
where $\Delta\Sigma(\mu^2)$ and $\Delta G(\mu^2)$ are net quark and gluon helicity 
defined as, 
\begin{eqnarray} 
\label{eq:sigma} 
\Delta\Sigma(\mu^2) 
\hskip -0.1in &=&  \hskip -0.1in 
\sum_{i=q,\bar{q}} 
\int_0^1 dx\, \Delta f_{i/p}(x,\mu^2) 
\equiv
\int_0^1 dx 
\left( \Delta
u+\Delta\bar{u}+ \Delta d+\Delta\bar{d}
+ \Delta s+\Delta\bar{s}\right)(x,\mu^2)\, ,
\nonumber \\[2mm]
\Delta G(\mu^2)
\hskip -0.1in &=& \hskip -0.1in 
\int_0^1 dx\, \Delta f_{g/p}(x,\mu^2)
\equiv 
\int_0^1 dx\, \Delta g(x,\mu^2)\, .
\end{eqnarray} 
\noindent
In Eq.~(\ref{eq:ssr}), the factor $1/2$ in the right-hand-side of the equation is the
spin of each quark and anti-quark; and $L_q(\mu^2)$ and $L_g(\mu^2)$ represent
the quark and gluon orbital angular momentum contribution, respectively.
The parton helicity distributions, $\Delta f$, are the key ingredients for
solving the proton spin puzzle.

The experiment of inclusive electron-proton deep inelastic scattering (DIS),
$\ell(k)+h(p)\to \ell'(k')+X$ with a large momentum transfer 
$Q \equiv \sqrt{-q^2}\equiv \sqrt{-(k-k')^2}\gg 1/$fm,
discovered the quarks inside the proton in the nineteen sixties at SLAC.  
The same experimental setting with longitudinally polarized electron and 
longitudinally polarized proton can access the helicity states 
of quarks and gluons of the polarized proton.  
With the approximation of one-photon exchange between the scattering lepton 
and the proton, neglecting the $Z$-boson exchange, 
the difference of the polarized DIS cross sections with the proton's spin
reversed measures the polarized DIS structure function, $g_1(x_B,Q^2)$, as  
\begin{equation} 
\label{eq:g1}
\frac{1}{2}\left[
\frac{\mathrm{d}^2\sigma^{\,\begin{subarray}{c} \rightarrow \\[-1.3mm]
\leftarrow
             \end{subarray}}}{\mathrm{d} x_B\,\mathrm{d} Q^2}-
\frac{\mathrm{d}^2\sigma^{\,\begin{subarray}{c} \rightarrow \\[-1.3mm]
\rightarrow
             \end{subarray}}}{\mathrm{d} x_B\,\mathrm{d} Q^2}\right] \
\simeq \ \frac{4\pi\,\alpha_{em}^2}{Q^4}\, y \left( 2 - y \right)
g_1(x_B,Q^2)\, ,
\end{equation} 
where the terms suppressed by $x_B^2(M_p^2/Q^2)$ with proton mass $M_p$ 
have been neglected, $x_B\equiv Q^2/2p\cdot q$ is the Bjorken variable, and
$y=q\cdot p/k\cdot p$ is the inelasticity of DIS.  In terms of QCD factorization \cite{CSS-FAC},
the DIS structure function $g_1$ can be factorized into a sum of parton helicity
distributions with perturbatively calculable coefficients expressed in a power series of 
strong coupling $\alpha_s$ 
\cite{Dokshitzer:1977sg,Gribov:1972ri,Altarelli:1977zs,Zijlstra:1993sh,Mertig:1995ny,Vogelsang:1995vh,Vogt:2008yw}.  At the leading order (LO) in $\alpha_s$, 
\begin{equation} 
\label{eq:g1plo}
g_1(x_B,Q^2) = \frac{1}{2} \sum e_q^2 \left[
\Delta q(x_B,Q^2) + \Delta \bar{q}(x_B,Q^2) \right] ,
\end{equation} 
where $e_q$ denotes a quark's electric charge, and the resolution scale
of the parton helicity distributions is set to be equal to the resolution scale of the exchange
virtual photon, $\mu=Q$.   It is clear from Eq.~(\ref{eq:g1plo}) that 
the structure function $g_1(x_B,Q^2)$ is directly sensitive to the
proton's spin structure in terms of the {\it combined} quark and anti-quark helicity distributions. 
Because of the electroweak probe of the inclusive DIS measurements, 
the gluon helicity distribution $\Delta g$ enters the
expression for $g_1$ only at higher order in perturbation theory.  
Since $\Delta g$ also contributes to the scaling violations 
(the resolution scale $\mu^2$-dependence) of quark and anti-quark 
helicity distributions, the access to gluon polarization by DIS measurements of $g_1$ 
requires a large level arm in $Q^2$, which could be achieved 
at a future Electron-Ion Collider (EIC) \cite{Accardi:2012qut}.

With the various hadronic probes, such as jet(s) and identified hadron(s), and 
electroweak probes, such as the photon, lepton(s), and $W/Z$-bosons, 
the RHIC spin program and its measurements of many complementary observables 
provide a much more direct access to gluon polarization, 
as well as flavor separation of quark polarizations.   

%======================================================================
\subsection{The gluon helicity structure}
\label{subsec:gluon}
%======================================================================

Longitudinally polarized proton-proton collisions at RHIC 
allow access to gluon helicity distribution, $\Delta g(x,\mu^2)$, 
at LO in perturbative QCD (pQCD).  With its detector advantages,
the PHENIX experiment makes the best connection to $\Delta g$ 
via single inclusive $\pi^0$ production at large transverse momentum $p_\perp$
by measuring the inclusive double-helicity asymmetries, $A_{LL}$, 
defined as\begin{equation}
A_{LL}=\frac{\Delta\sigma}{\sigma}
=\frac{\sigma_{++}-\sigma_{+-}}{\sigma_{++}+\sigma_{+-}}\, ,
\label{eq:all}
\end{equation}
where $\Delta\sigma$ ($\sigma$) is the polarized (unpolarized) cross section,
and $\sigma_{++}$ ($\sigma_{+-}$) represents the cross section of $\vec{p}+\vec{p}$ 
collisions with the same (opposite) proton helicity.  In terms of QCD factorization, both the 
polarized and unpolarized cross sections for the single inclusive $\pi^0$ production can
be factorized to show the explicit connection to the parton helicity distributions and PDFs, 
respectively,
\begin{equation}
A_{LL}^{\pi^0}
=
\frac{
\sum_{abc} \Delta f_{a/p}(x_1) \otimes \Delta f_{b/p}(x_2) 
\otimes \Delta\hat{\sigma}_{a+b\to c+X}(x_1,x_2,p_c) \otimes D_{c\to\pi^0}(z)
}{
\sum_{abc} f_{a/p}(x_1) \otimes f_{b/p}(x_2) \otimes 
\hat{\sigma}_{a+b\to c+X}(x_1,x_2,p_c) \otimes D_{c\to\pi^0}(z)
} ,
\label{eq:all-fac}
\end{equation}
where $x_1$ and $x_2$ are momentum fractions of two colliding partons, 
$D_{c\to\pi^0}(z)$ is the fragmentation function (FF) representing a probability distribution 
for a parton $c$ of momentum $p_c$ to fragment into a hadron $\pi^0$ of momentum $z\,p_c$,
$\otimes$ represents the convolution over the active parton's momentum fractions, 
and the dependence on the factorization and renormalization scales are suppressed.
In Eq.~(\ref{eq:all-fac}), $\Delta\hat{\sigma}_{a+b\to c+X}$ ($\hat{\sigma}_{a+b\to c+X}$) is the 
polarized (unpolarized) scattering cross section of a partonic subprocess, $a+b\to c+X$, 
for two active incoming partons of flavor $a$ and $b$, respectively, 
to produce a single inclusive parton of flavor $c$ and momentum $p_c$.  
Both polarized and unpolarized partonic cross sections are calculable in pQCD, and 
are available for LO as well as next-to-leading order (NLO) in powers of $\alpha_s$. 
For the kinematic region covered by the RHIC energies, 
partonic cross sections of gluon initiated subprocesses 
are significantly larger than those initiated by quarks and antiquarks.
That is, measurements of the asymmetry $A_{LL}^{\pi^0}$ by PHENIX Experiment 
at RHIC energies could provide very sensitive information on the gluon helicity distribution. 

The factorized expression for $A_{LL}^{\pi^0}$ in Eq.~(\ref{eq:all-fac}) should be valid for 
all single inclusive hadron production so long as the produced hadron 
transverse momentum is much larger than its mass, $p_{h\perp}\gg M_h$, to ensure the validity 
of QCD factorization, and the $\pi^0$ FFs, $D_{c\to\pi^0}(z)$, are replaced by 
corresponding hadron FFs, $D_{c\to h}(z)$.  This is because all factorized 
partonic scattering cross sections are insensitive to the details of hadronic states produced.
Having the universal FFs, $D_{c\to h}(z)$, extracted from other scattering processes, 
such as $e^+ + e^-\to h(p_h)+X$, measurements of $A_{LL}^{h}$ for the production of 
single inclusive hadron $h$ other than $\pi^0$ provide additional information on 
the universal parton helicity distributions.  The PHENIX Experiment measured the 
asymmetry $A_{LL}^{\pi^0}$ 
\cite{Adler:2004ps,Adare:2007dg,Adare:2008px,Adare:2008qb,Adare:2014hsq}, 
as well as $A_{LL}^{\eta}$ \cite{Adare:2010cy}.  

By removing the dependence on the FFs, and replacing the single-parton inclusive cross sections,
$\Delta\hat{\sigma}_{a+b\to c+X}$ and $\hat{\sigma}_{a+b\to c+X}$ 
by corresponding partonic jet cross sections, $\Delta\hat{\sigma}_{a+b\to \mbox{jet} +X}$ and 
$\hat{\sigma}_{a+b\to \mbox{jet} +X}$, respectively, 
the factorized expression for $A_{LL}^{\pi^0}$ in Eq.~(\ref{eq:all-fac}) reduces to
the factorized expression for $A_{LL}^{\rm jet}$, which depends on the same parton helicity 
distributions and PDFs.  Precise measurements of the double longitudinal spin asymmetry 
for jet production at RHIC energies could also provide 
excellent information on parton helicity distributions
\cite{Abelev:2006uq,Abelev:2007vt,Sarsour:2009zd,Djawotho:2011zz,Adamczyk:2014ozi}.

With the earlier measurements of the double helicity asymmetry, $A_{LL}^{\pi^0}$, by the PHENIX Experiment at the mid-rapidity for $\sqrt{s}=200$~GeV collision energy in 2005 and 2006 
\cite{Adler:2004ps,Adare:2007dg,Adare:2008px,Abelev:2009pb,Adare:2008qb},
along with measurements of $A_{LL}^{\rm jet}$ by the STAR Experiment 
\cite{Abelev:2006uq,Abelev:2007vt,Sarsour:2009zd,Djawotho:2011zz},
the NLO QCD global analysis, known as the ``DSSV" analysis, concluded that
the RHIC data -- within their uncertainties at that time -- 
did not show any evidence of a net polarization of gluons inside the proton 
\cite{deFlorian:2008mr,deFlorian:2009vb}.

Recently, based on data collected in 2009 at $\sqrt{s}=200$~GeV at RHIC, 
which not only approximately doubles the statistics of the earlier measurements
\cite{Adare:2008px,Adare:2010cy}, but also extends the range of measured $p_\perp$,
PHENIX Collaboration published new measurements for both $A_{LL}^{\pi^0}$ 
and $A_{LL}^{\eta}$ \cite{Adare:2014hsq},
%%%%%%%%%%%%%%%%%%%%%%%%%%%%%%%%%%%%%%%%%%%%%%%%
\begin{figure}[!h]
\begin{center}
\vspace{-0.1in}
\includegraphics[width=0.45\textwidth]{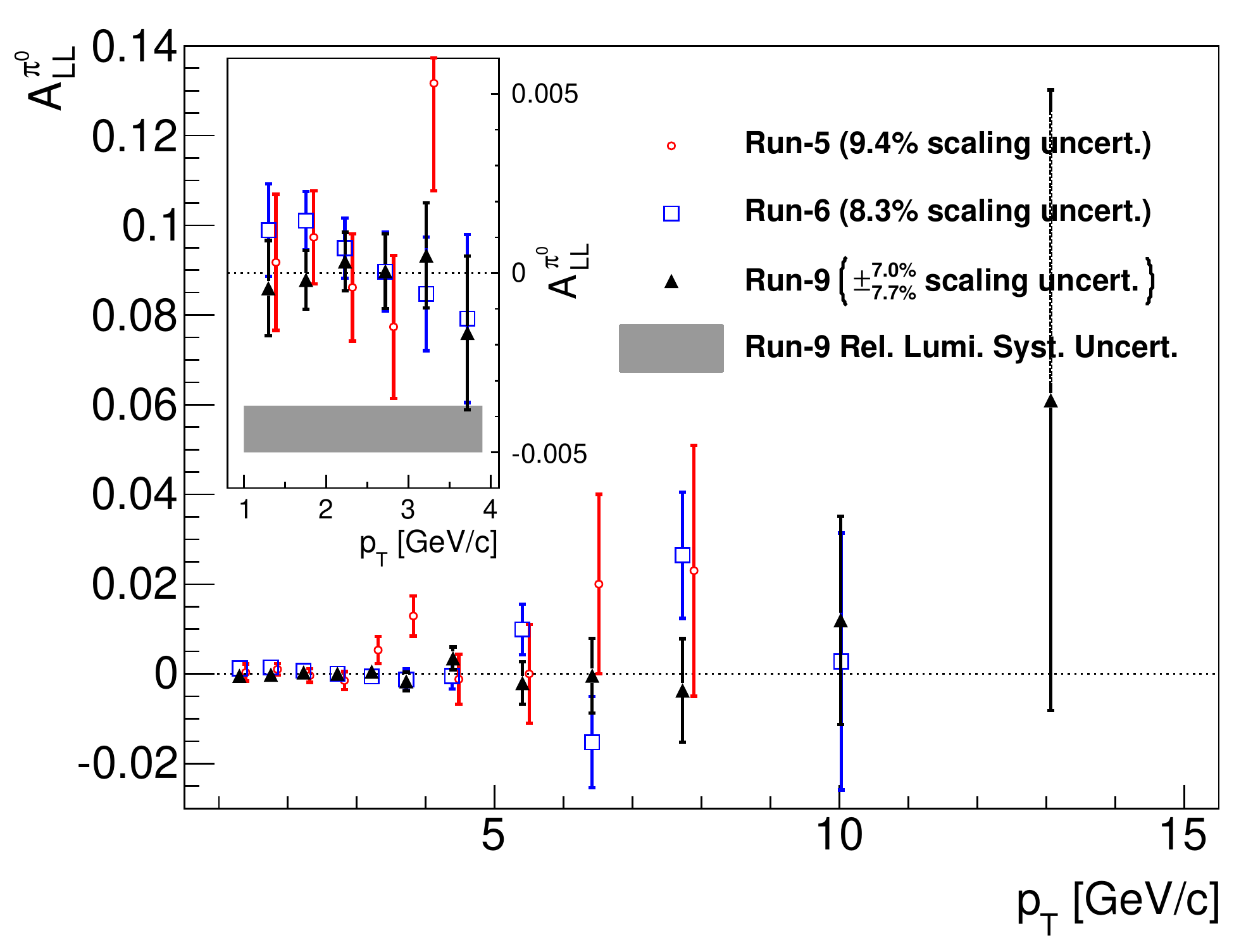}
\hskip 0.1in
\includegraphics[width=0.45\textwidth]{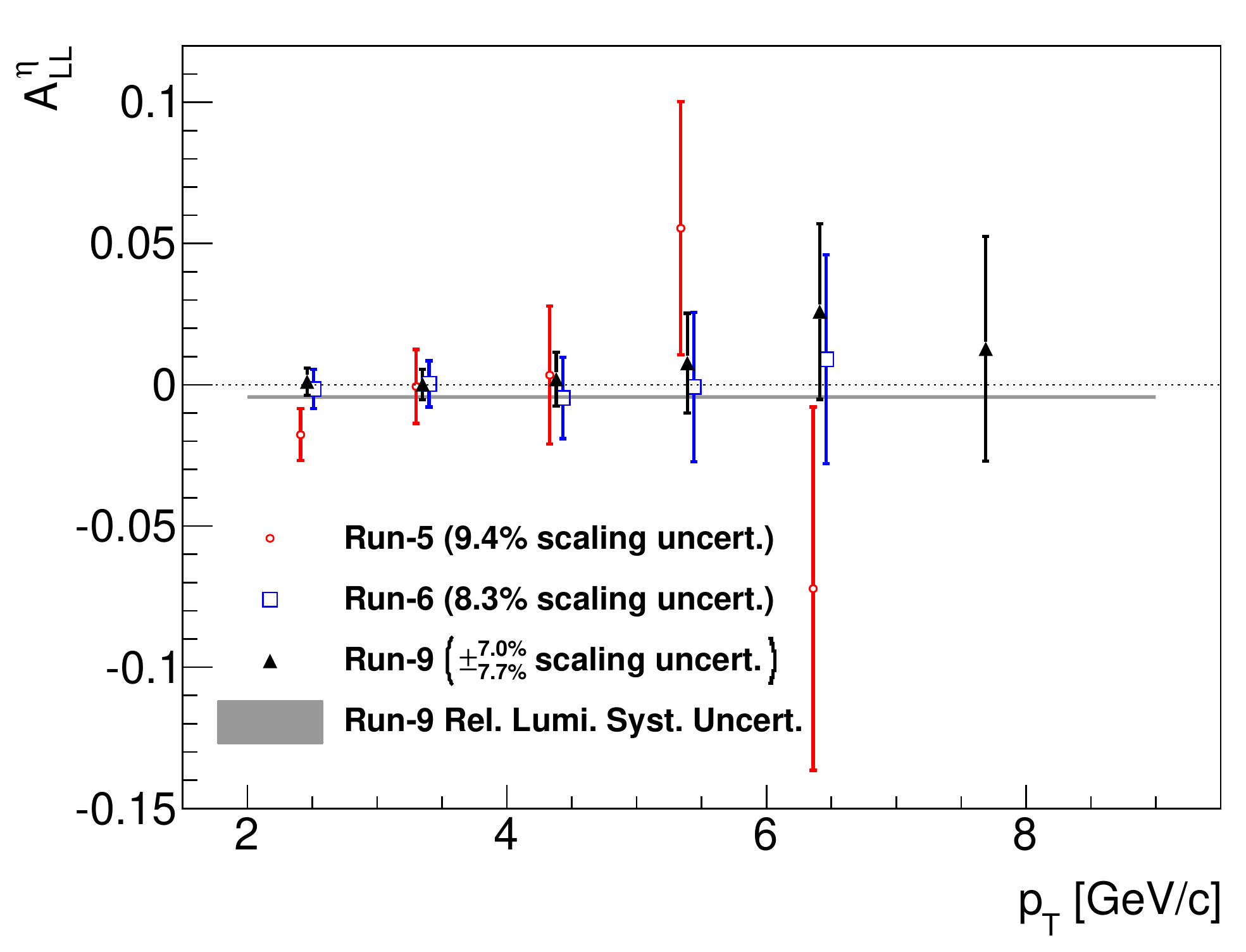}
\caption{$A_{LL}^{\pi^0}$ (Left) and $A_{LL}^{\eta}$ (Right) as a function 
of meson's transverse momentum $p_T$ for the 2005 (red circle), 
2006 (blue square) and 2009 (black triangle) PHENIX data sets.  
\label{fig:pi0-eta}}
\vspace{-0.2in}
\end{center}
\end{figure}
As shown in Fig.~\ref{fig:pi0-eta}, three data sets collected during the 2005, 2006, and 2009
RHIC runs with polarized proton collisions are consistent with each other. 

With the better statistics and extended $p_\perp$ range, new data from 2009 RHIC run 
could naturally put tighter constraints on gluon helicity distribution, $\Delta g(x)$, and extend
the range of $x$ over which meaningful constraints can be obtained from QCD global fits.
Within experimental uncertainties, the new PHENIX data on $A_{LL}^{\pi^0}$ and $A_{LL}^{\eta}$ 
are still consistent with the existing DSSV analysis, and other recent NLO QCD global 
analyses of DIS-only data by Bl{\"u}mlein and B{\"o}ttcher (BB10)~\cite{Blumlein:2010rn} 
and Ball et.~al.~(NNPDF)~\cite{Ball:2013lla,Ball:2012cx},
as well as the analysis by Leader et. al. (LSS10)~\cite{Leader:2010rb} based on 
both DIS and Semi-Inclusive DIS (SIDIS) data.  Since various analyses used different 
assumptions on the symmetry properties of parton helicity distributions, and different 
functional forms for the distributions at the input scale, the determination of the 
gluon helicity distribution, $\Delta g(x,\mu^2)$ varies, see the detailed discussion in the 
PHENIX publication \cite{Adare:2014hsq}.

\begin{figure}[!h]
\begin{center}
\vspace{-0.1in}
\includegraphics[width=0.51\textwidth]{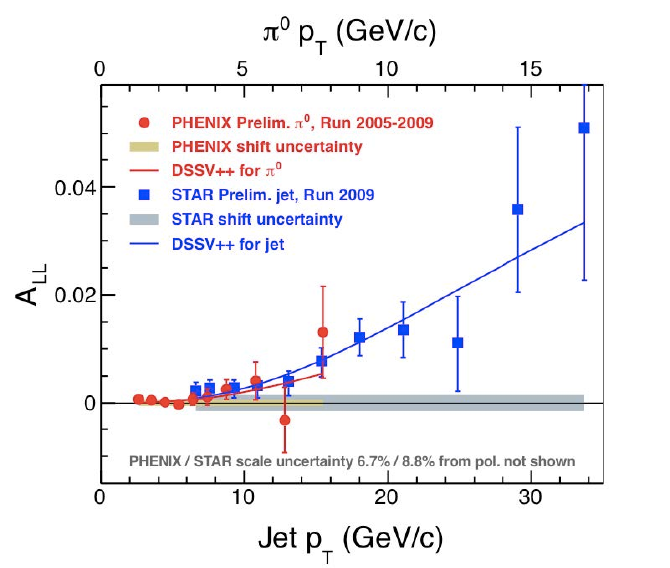}
\includegraphics[width=0.46\textwidth,height=2.5in]{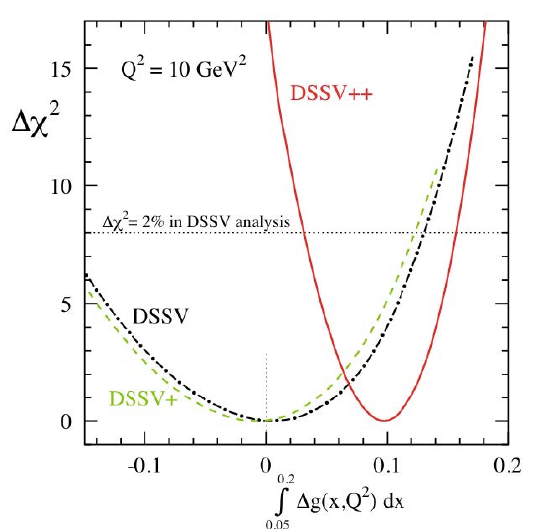}
\caption{Left:\ Double-spin asymmetry $A_{LL}$ of new PHENIX and STAR data are 
compared with the new DSSV{\small{++}} QCD global analysis \cite{deFlorian:2014yva}. 
Right:\ the $\chi^2$ profile for the integrated gluon helicity contribution in
the $x$ region currently probed at RHIC for $\sqrt{s}= 200$~GeV.
\label{fig:new-dssv}}
\vspace{-0.2in}
\end{center}
\end{figure}

Although the new PHENIX data on $A_{LL}^{\pi^0}$ and $A_{LL}^{\eta}$ 
do not show any significant asymmetry \cite{Adare:2014hsq}, 
the new STAR data on $A_{LL}^{\rm jet}$ of inclusive jet production 
from the same 2009 run at RHIC
show a non-vanished double-spin asymmetry 
over the whole range $5\lesssim p_T\lesssim 30$~GeV \cite{Adamczyk:2014ozi}, 
which differs from the previous results on jet production published 
by the STAR Collaboration \cite{Abelev:2006uq,Abelev:2007vt,Sarsour:2009zd}.
The most resent DSSV QCD global analysis, referred as ``DSSV{\small{++}}'' 
\cite{deFlorian:2014yva}, shows that the new PHENIX data and STAR data
are actually consistent with each other, as shown in Fig.~\ref{fig:new-dssv}(Left).  
The new data sets and new DSSV{\small{++}} QCD global analysis lead to a significant 
net positive gluon helicity contribution from the $x$ region probed by the current
measurements at RHIC, $0.05\lesssim x\lesssim 0.2$, with a better constrained 
$\chi^2$ profile for the global analysis, as shown in Fig.~\ref{fig:new-dssv}(Right).
For the first time, we have a clear experimental evidence to show that gluon
could give a non-vanish contribution to the proton's spin, at least from the 
momentum region probed by the current RHIC measurements.  
%%%%%%%%%%%%%%%%%%%%%%%%%%%%%%%%%%%%%%%%%%%%%%%%

With new measurements of $\pi^0$ in large forward rapidity region 
at $\sqrt{s}=500$~GeV, a higher center-of-mass energy of polarized proton-proton 
collisions at RHIC, the PHENIX Experiment with its forward calorimeter can access 
the gluon helicity distribution at a momentum fraction $x$ as small as 0.002, which 
will considerably improve our knowledge of the gluon spin contribution to 
the proton's spin in the $x$-range of $0.002-0.2$~\cite{Aschenauer:2013woa}.

%======================================================================
\subsection{The sea quark helicity structure}
\label{subsec:quark}
%======================================================================

The combination of the large body of inclusive DIS data has provided excellent 
measurements of combined quark and antiquark helicity structure, and established 
that the up quarks and anti-upquarks combine to have net
polarization along the proton spin, whereas the down quarks and anti-downquarks 
combine to carry negative polarization.  The ``total''
$\Delta u+\Delta\bar{u}$ and $\Delta d+\Delta\bar{d}$ helicity
distributions are very well constrained for the medium to large $x$,
contributing to about 30\% of the proton's spin.

However, the light sea quark and anti-quark structure 
is still far from being well-understood \cite{Peng:2014hta}, 
and its helicity distributions still carry large uncertainties, 
even though there are some constraints by the SIDIS data.
Taking the advantage of the fact that $W$ bosons couple only 
the left-handed quarks and right-handed antiquarks 
($u_L\bar{d}_R\to W^+$ and $d_L \bar{u}_R\to W^-$), 
measuring the single longitudinal spin asymmetry, 
$A_L=(\sigma_+-\sigma_-)/(\sigma_++\sigma_-)$, 
of the parity violating production of $W$-bosons from
flipping the helicity of one of the polarized colliding protons
at RHIC can probe the flavor dependence of sea quark helicity 
distributions, $\Delta q$ and $\Delta\bar{q}$. 
%%%%%%%%%%%%%%%%%%%%%%%%%%%%%%%%%%%%%%%%%%%%%%%%
\begin{figure}[!h]
\begin{center}
\vspace{-0.1in}
\includegraphics[width=0.7\textwidth]{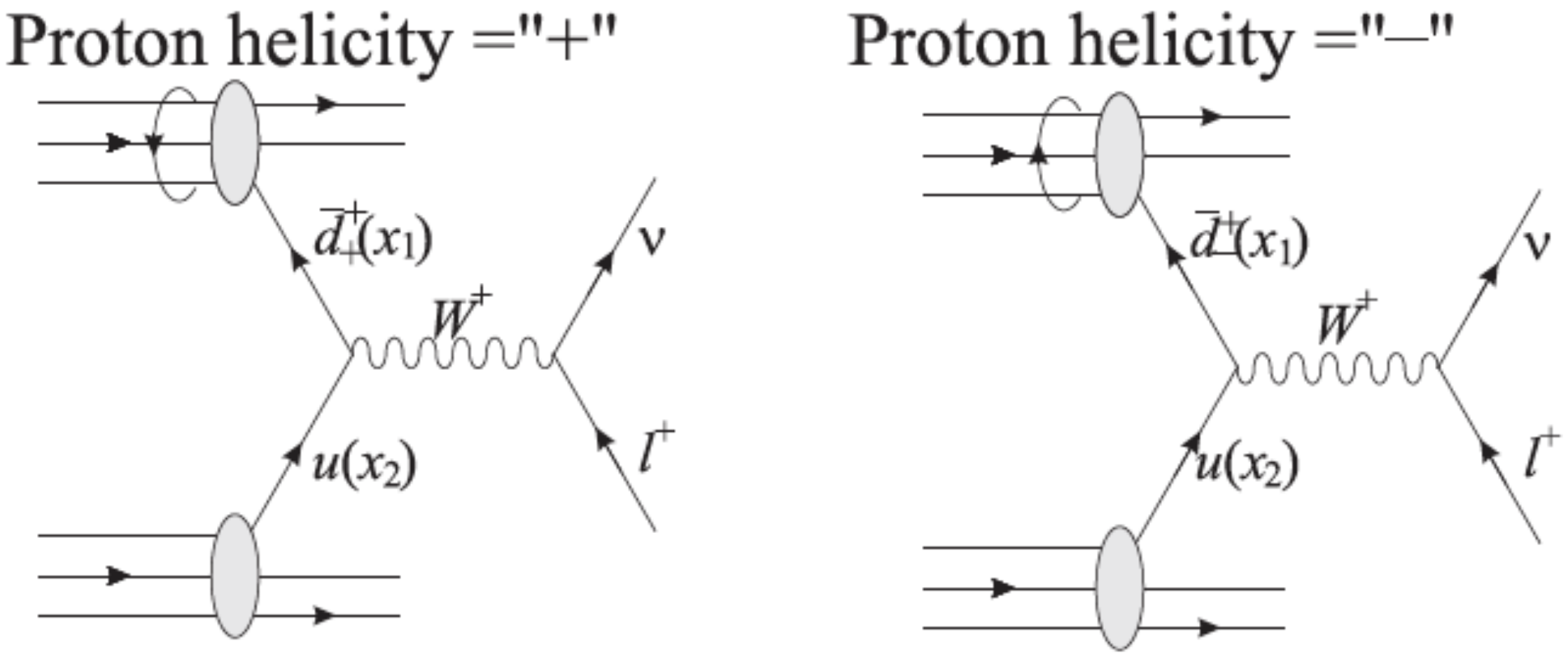}
\caption{Sketch for $W^+$ production from a leading order 
annihilation of a right-handed anti-downquark and a left-handed 
up quark in the collision of a polarized proton (top) and a unpolarized proton (bottom).  
The produced $W^+$ then decays to a positron and an electron neutrino.
\label{fig:w-prod}}
\vspace{-0.2in}
\end{center}
\end{figure}
For example, the difference of the $W^+$ yield from the diagram
on the left in Fig.~\ref{fig:w-prod} and that from the diagram on the right
is of good information on anti-downquark helicity distribution, 
$\Delta\bar{d}(x_1) = \bar{d}_+ ^+(x_1) - \bar{d}_-^+(x_1)$, where
the subscript $+$ ($-$) indicates the positive (negative) helicity state of 
the polarized proton, the superscript $+$ represents the handiness of 
the anti-downquark, and $x_1$ is the momentum fraction of the polarized
proton carried by the anti-downquark.
Production of the $W$-boson requires a large momentum
transfer at a scale where higher order QCD corrections can
be evaluated reliably, and is free from uncertainties in 
fragmentation functions that are needed for probing light hadrons.
The measured $W$-boson cross sections in unpolarized hadronic
collisions at Tevatron and the LHC, as well as at RHIC, 
have confirmed our theoretical understanding of the production mechanism.

The measurements of $A_L$ of $W^\pm$-bosons at RHIC as a function
of decay lepton's transverse momentum and rapidity provide the uniquely
clean information on the sea quark helicity distributions, as well as its flavor structure 
inside a polarized fast moving proton.  With the $W$ mass setting up the hard scale for the 
momentum transfer of the collisions, RHIC's $W$-physics program could
probe the quark and anti-quark helicity distributions at a medium 
momentum fraction, $0.05 \lesssim x \lesssim 0.4$, providing 
strong constraints on the sea quark contribution to the proton's spin.
Even more important, RHIC's $W$-physics program could provide 
the unique and much needed information on the sea quark helicity 
asymmetry, such as $\Delta\bar{d}(x)-\Delta\bar{u}(x)$, in a perfect kinematic region,
in view of the puzzling sign change of unpolarized sea quark distributions, 
$\bar{d}(x)-\bar{u}(x)$, at $x\sim 0.3$ observed by the E886/NuSea
Collaboration at Fermilab in its extraction of $\bar{d}(x)-\bar{u}(x)$
over the region $0.02 < x < 0.345$ from the measurements of 
Drell-Yan process \cite{Hawker:1998ty,Peng:1998pa,Towell:2001nh}.  
The sea quark asymmetry in both momentum and helicity distributions 
is fundamentally important for understanding the proton structure.

From the background subtracted yields of positions and electrons in $p$-$p$ collisions 
at $\sqrt{s}=500$~GeV at RHIC, PHENIX experiment extracted the 
$W^+$ and $W^-$ boson production cross sections, 
$\sigma(pp\!\rightarrow\!W^{+}X)\times BR(W^+\!\rightarrow\!e^+\nu_e)=144.1\pm 
21.2 ({\rm stat})^{+3.4}_{-10.3} ({\rm syst})\pm15\% (\rm norm)$pb,
and $\sigma(pp\!\rightarrow\!W^{-}X)\times 
BR(W^-\!\rightarrow\!e^-\bar{\nu_e})=31.7\pm 
12.1 ({\rm stat})^{+10.1}_{-8.2} ({\rm syst})\pm15\% (\rm norm)$pb, respectively,
where $BR$ is the $W$'s decay branching ratio \cite{Adare:2010xa}.  
The measurements of PHENIX Collaboration, 
together with that of STAR Collaboration \cite{Corliss:2014dha}, 
provided in fact the very first $W$-boson production cross sections in 
$p$-$p$ collisions, which are consistent with earlier measurements 
in $\bar{p}$-$p$ collisions at CERN and Fermilab \cite{Adare:2010xa}.
%%%%%%%%%%%%%%%%%%%%%%%%%%%%%%%%%%%%%%%%%%%%%%%%
\begin{figure}[!t]
\begin{center}
\includegraphics[width=0.57\textwidth,height=2.4in]{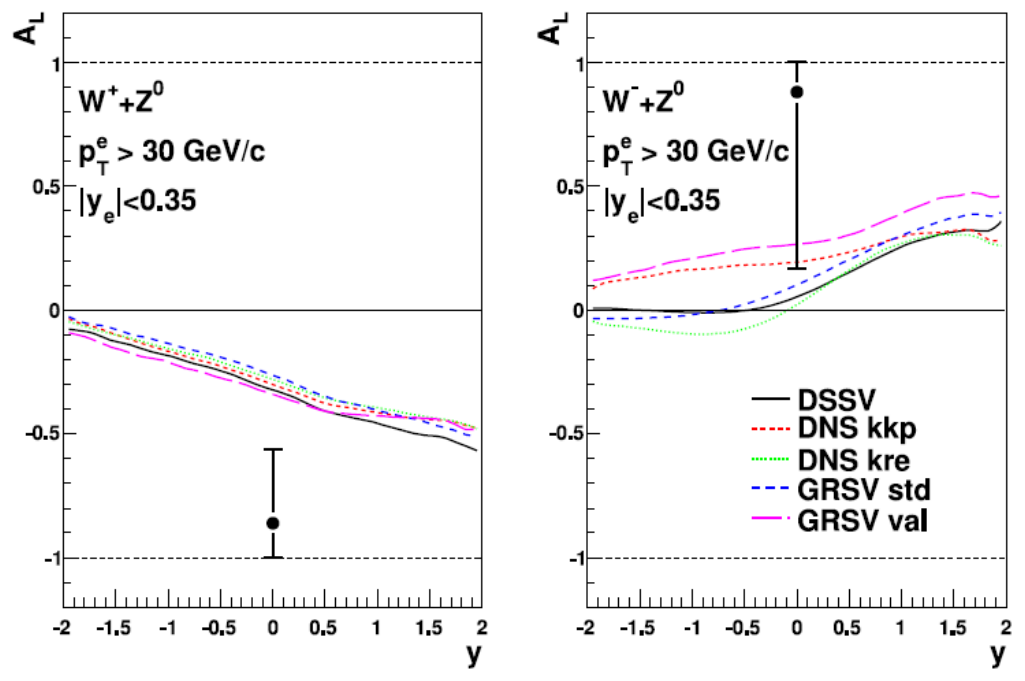}
\hskip 0.1in
\includegraphics[width=0.38\textwidth,height=2.4in]{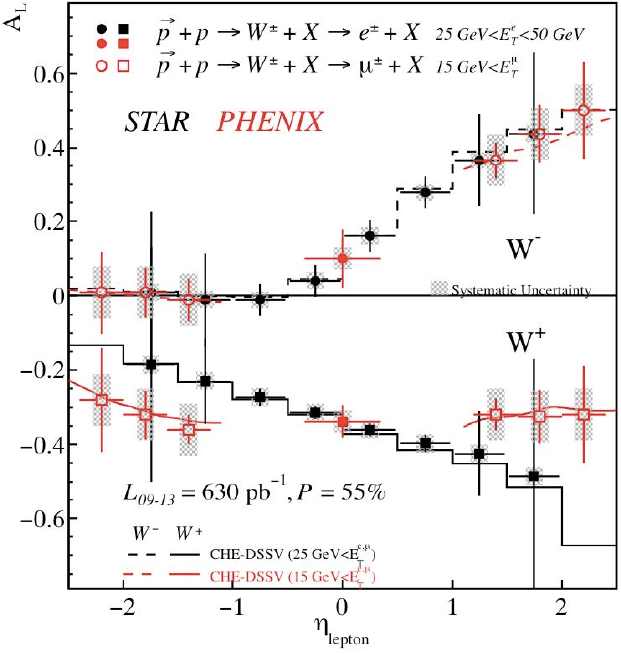}
\caption{PHENIX measurement of $A_L$ for 
electrons (Left) and positions (Middle) from $W$ and $Z$ decays \cite{Adare:2010xa}, along with
theory predictions \cite{deFlorian:2010aa}.
Right:\ Expected uncertainties for $A_L$ of $W^{\pm}$ production for PHENIX and STAR after
the 2013 run. The asymmetries have been randomized around the central value of DSSV analysis.
\label{fig:phenix-al}}
\vspace{-0.3in}
\end{center}
\end{figure}

With its early data on leptonic $W$ decay
in polarized $p$-$p$ collisions at $\sqrt{s} = 500\,\mathrm{GeV}$,
RHIC's $W$-physics program started to impact our understanding of 
proton's polarized sea structure \cite{Adamczyk:2014xyw,Gal:2014fha}.  
In Fig.~\ref{fig:phenix-al}, the very first PHENIX measurement of the parity 
violating single spin asymmetry, $A_L$, of electrons (Left) and positions (Right) 
from $W$ and $Z$ decays are presented \cite{Adare:2010xa}.  
Although the experimental uncertainties are clearly large, 
the asymmetries are consistent with the early theory expectations. 
%%%%%%%%%%%%%%%%%%%%%%%%%%%%%%%%%%%%%%%%%%%%%%%%
\begin{figure}[!b]
\begin{center}
\vspace{-0.1in}
\includegraphics[width=0.45\textwidth]{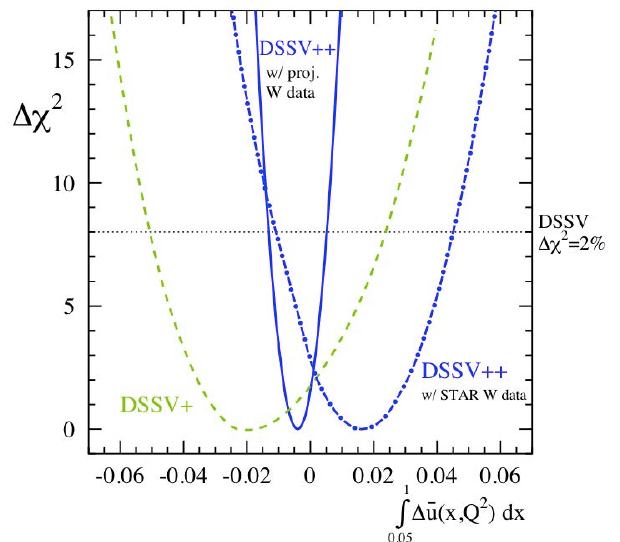}
\hskip 0.1in
\includegraphics[width=0.45\textwidth]{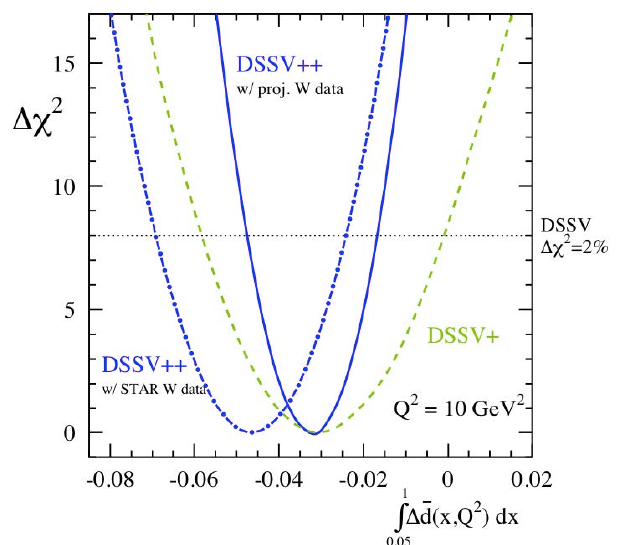}
\caption{Expected $\chi^2$ profiles from DSSV global analyses of parton
helicity distributions for $\Delta\bar{u}$ (left) and $\Delta\bar{d}$ (right):  DSSV{\small{+}}
with new SIDIS data, and DSSV{\small{++}} with new STAR data or with projected data 
on $A_L$ of $W^\pm$ boson production in Fig.~\ref{fig:phenix-al} (Right). 
\label{fig:projected-sea}}
\vspace{-0.2in}
\end{center}
\end{figure}

With the new STAR data on $A_L^{W^\pm}$ \cite{Adamczyk:2014xyw}, 
and the expected experimental uncertainties for $A_L$ measurements of 
both PHENIX and STAR from the recent polarized $p$-$p$ runs at RHIC 
\cite{Aschenauer:2013woa}, as shown in Fig.~\ref{fig:phenix-al} (Right), 
new and updated DSSV global analyses of parton helicity distributions clearly indicate
that the sea quark structure shows hints of not being SU(2)-flavor symmetric: 
the $\Delta \bar{u}$ distribution has a tendency to be mainly positive, 
and the $\Delta \bar{d}$ anti-quarks likely carry an opposite polarization. 
From the $\chi^2$ profiles of the global fits in Fig.~\ref{fig:projected-sea}, 
the new RHIC data on the parity violating single spin asymmetries will provide 
a powerful constraint on the sea quark structure and its contribution to the proton's spin.

QCD global analyses of all available data from DIS, SIDIS, and RHIC have provided 
good information on quark/gluon helicity structure of a polarized fast moving proton, 
and quark/gluon spin contribution to the proton's spin \cite{Accardi:2012qut}.  
From the region of momentum fractions accessible by existing experimental measurements, 
the net helicity of quarks and antiquarks combined could account for about 30\% of proton's spin,
while gluons are likely to give a positive contribution to the proton's spin, close to 20\% 
from the most recent RHIC data, but with a hugh uncertainty from the region 
of gluon momentum fractions that is not available to all existing and near future facilities 
\cite{deFlorian:2014yva}.  While we need the final word of the gluon spin contribution
to the proton's spin from the proposed future EIC \cite{Accardi:2012qut}, it is 
fundamentally important to investigate the confined transverse motion of quarks 
and gluons inside a fast moving proton, and their contribution to the proton's spin.

%%%%%%%%%%%%%%%%%%%%%%%%%%%%%%%%%%%%%%%%%%%%%%%%
\section{Transverse parton structure of the proton }
\label{sec:tmds}
%%%%%%%%%%%%%%%%%%%%%%%%%%%%%%%%%%%%%%%%%%%%%%%%

To have a direct access to the transverse parton structure of a fast moving proton 
in high energy scattering requires a new type of physical observables 
-- the controlled probes that have at least two distinctive momentum 
scales: a large scale $Q \gg \Lambda_{\rm QCD}\sim 1/\text{fm}$ to localize the probes to 
``see'' the particle nature of quarks and gluons, and a small scale $q\ll Q$ to be sensitive to 
the confined transverse motion or spatial distribution of quarks and gluons inside the proton.  

Theoretically, the simplest transverse parton structure of the proton could be characterized 
by the Wigner distributions $W_{i/p}(x, \boldsymbol{k}_T, \boldsymbol{b}_T)$ -- 
the probability distributions to find a parton of flavor $i (=q,\bar{q},g)$ 
inside the fast moving proton, carrying its momentum fraction between $x$ and $x+dx$
and having a transverse momentum $\boldsymbol{k}_T$ at a transverse spatial position
$\boldsymbol{b}_T$ with respect to its center \cite{Belitsky:2003nz}, 
More complex transverse parton structure of the proton could be represented 
by multi-parton correlation functions in terms of longitudinal momentum fractions, 
transverse momenta and positions of all active partons.  
Although the Wigner distributions are used extensively in other branches of physics 
\cite{Hillery:1983ms}, there has not been any known way to measure the quark and gluon 
Wigner distributions in high energy experiments.  

However, various reductions of the Wigner distributions could be probed in high energy 
experiments \cite{Accardi:2012qut}.  By integrating over $\boldsymbol{b}_T$, 
the Wigner distributions are reduced to the three-dimensional (3D) momentum distributions 
of quarks and gluons inside the fast moving proton 
-- transverse momentum dependent PDFs (TMDs), 
$f_{i/p}(x,\boldsymbol{k}_T)$; and the TMDs could be further reduced to 
the better known PDFs, $f_{i/p}(x,\mu^2)$ by integrating over $\boldsymbol{k}_T$ 
with a proper ultraviolet (UV) renormalization when $\boldsymbol{k}_T\to\infty$
for the quark/gluon operators defining the PDFs at the renormalization scale $\mu$.
Similarly, by integrating over $\boldsymbol{k}_T$, the Wigner distributions are reduced to 
the impact parameter distributions, $f_{i/p}(x,\boldsymbol{b}_T)$, 
the probability distributions to find a parton of flavor $i$ and momentum fraction $x$ 
at a transverse spatial position $\boldsymbol{b}_T$, 
which are effectively the tomographic images of quarks and gluons inside the proton.  
Furthermore, by taking the Fourier transform of 
$\boldsymbol{b}_T$ into the conjugate momentum $\boldsymbol{\Delta}_T$, 
$f_{i/p}(x,\boldsymbol{b}_T)$ could be transformed into the generalized PDFs 
(GPDs), such as $H_{i/p}(x,\xi,t)$ and $E_{i/p}(x,\xi,t)$ at $\xi=0$ 
with $t=-\boldsymbol{\Delta}_T^2$ \cite{Mueller:2014hsa}.  
The GPDs could be extracted from measurements of exclusive processes 
in lepton-hadron DIS or in ultra-peripheral hadronic collisions.  
The TMDs and GPDs represent various aspects of the same proton's 
transverse structure of quarks and gluons that could be probed in high energy scattering
\cite{Accardi:2012qut}.  In fact, some GPDs are intimately connected with the
orbital angular momentum carried by quarks and gluons \cite{Burkardt:2005km}.  
The Ji's sum rule is one of the examples that quantify this connection \cite{Ji:1996ek}, 
\begin{eqnarray}
J_q = \frac{1}{2} \lim_{t\to 0} \int_0^1 dx\, x \left[
H_q(x,\xi,t) + E_q(x,\xi,t) \right]\, ,
\label{eq:jissumrule}
\end{eqnarray}
which represents the total angular momentum $J_q$ (including both helicity and orbital
contributions) carried by quarks and anti-quarks of flavor $q$.  A similar relation holds 
for gluons.  

With the polarized high energy proton beam(s), the RHIC spin program could access the 
proton's transverse structure of quarks and gluons by measuring various TMDs, 
as well as GPDs via exclusive processes in ultra-peripheral 
${p}$-$p$ and ${p}$-$A$ collisions. 
The RHIC spin program could measure many emergent QCD phenomena, such as 
quantum correlations between the proton's spin direction and preference of the 
confined transverse motion of quarks and gluons inside the polarized proton, 
known as the Sivers effect \cite{Sivers:1989cc}.  Furthermore, it could probe
quantum correlation between the spin direction of the fragmenting parton and 
preference in direction where the produced hadron emerges, known as the
Collins effect \cite{Collins:1992kk}.  

%======================================================================
\subsection{Transverse single-spin asymmetry -- $A_N$}
\label{subsec:an}
%======================================================================

\begin{figure}[!h]
\begin{center}
\vspace{-0.1in}
\includegraphics[width=0.97\textwidth]{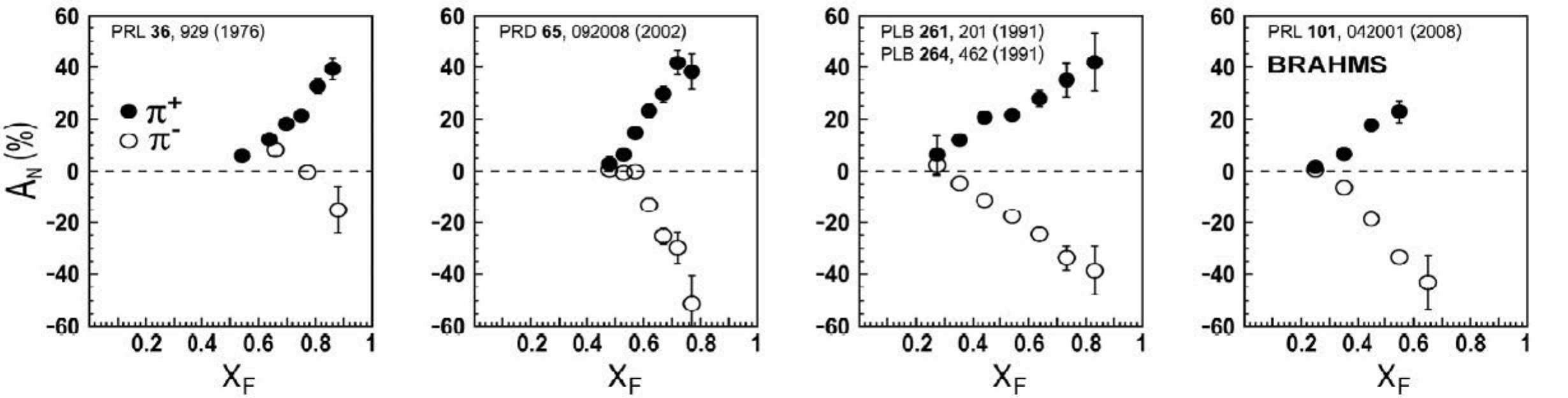}
\vspace{-0.1in}
\caption{Persistent transverse single-spin asymmetry, $A_N$, in hadronic 
single inclusive pion production.
\label{fig:an-pion}}
\vspace{-0.2in}
\end{center}
\end{figure}

Transverse single-spin asymmetry (SSA),
$A_N \equiv (\sigma(s_T)-\sigma(-s_T))/(\sigma(s_T)+\sigma(-s_T))$, 
is defined as the ratio of the difference and the sum of the 
cross sections when the spin of one of the identified hadron $s_T$ is flipped. 
Large SSAs of inclusive single pion production with a large momentum transfer 
in hadronic collisions have been consistently observed since nineteen seventies,  
as shown in Fig.~\ref{fig:an-pion}, although the asymmetries were once thought 
impossible in QCD \cite{Kane:1978nd}.  
With over two decades of intense theoretical as well as experimental efforts, 
our understanding of the observed large SSAs in QCD has been much improved.
Large SSAs are not only possible in QCD, but also carry extremely valuable information 
on the transverse motion and structure of quarks and gluons 
inside a transversely polarized proton.   
%%%%%%%%%%%%%%%%%%%%%%%%%%%%%%%%%%%%%%%%%%%%%%%%

From the parity and time-reversal invariance of QCD dynamics, non-vanish transverse
SSAs are necessarily connected to the transverse momentum of quarks and gluons inside
the transversely polarized proton \cite{Qiu:1990cu}.
Two complementary QCD-based approaches have been proposed 
to describe the physics behind the measured SSAs: the TMD factorization approach
\cite{Sivers:1989cc,Collins:1992kk,Brodsky:2002cx,Collins:2002kn} 
and the collinear factorization approach 
\cite{Efremov:1981sh,Efremov:1984ip,Qiu:1991pp,Qiu:1991wg,Qiu:1998ia,Kouvaris:2006zy}. 
In the TMD factorization approach, the asymmetry is attributed to 
direct correlation between spin and transverse momentum, which are represented 
by the TMD PDFs or FFs, 
such as the Sivers and Collins functions, respectively. 
On the other hand, the asymmetry in the collinear factorization approach 
is generated by twist-3 collinear PDFs or FFs.  With the transverse momenta of all
active partons integrated, the explicit spin-transverse momentum correlation 
in the TMD approach is now replaced by the integrated net effect of 
spin-transverse momentum correlations, which are effectively generated 
by QCD color Lorentz force \cite{Qiu:1993af}.  

Two approaches each have their own kinematic domain of validity, 
while they are consistent with each other in the perturbative regime 
where they both apply \cite{Ji:2006ub,Ji:2006vf,Ji:2006br,Koike:2007dg,Bacchetta:2008xw}. 
The TMD factorization approach is more suitable for evaluating the SSAs 
of scattering processes with two very different momentum transfers, 
$Q\gg q \gtrsim \Lambda_{\rm QCD}$, while the collinear factorization approach 
is more relevant to the SSAs of scattering cross sections with all observed momentum 
transfers hard and comparable: $Q'\sim Q\gg \Lambda_{\rm QCD}$. 
In hadornic collisions, the single inclusive hadron production at high $p_T$ 
is better treated in the collinear factorization approach, 
while the Drell-Yan and $W/Z$ production at a low transverse momentum, $q_T\ll Q$, 
needs the TMD factorization approach. 

%======================================================================
\subsection{$A_N$ of single hadron production}
\label{subsec:pion}
%======================================================================

The inclusive single hadron production at large transverse momentum $p_T$ at RHIC 
has effectively one large momentum transfer at ${\cal O}(p_T)\gg \Lambda_{\rm QCD}$.  
Collinear factorization approach is more suited for studying the transverse SSAs 
of the inclusive single hadron production.  In terms of the collinear factorization, 
the transverse SSAs are effectively power suppressed observables comparing 
to the production cross sections.  That is, the asymmetries are in general small, 
except in the region of phase space where the momentum spectrum of the observed 
pions is very steep, and a small shift in the spectrum could make the difference of two
cross sections with the proton spin flipped to be comparable with the cross section
itself, leading to a significant value of the asymmetry.  
%%%%%%%%%%%%%%%%%%%%%%%%%%%%%%%%%%%%%%%%%%%%%%%%
\begin{figure}[!h]
\begin{center}
\vspace{-0.1in}
\includegraphics[width=0.45\textwidth]{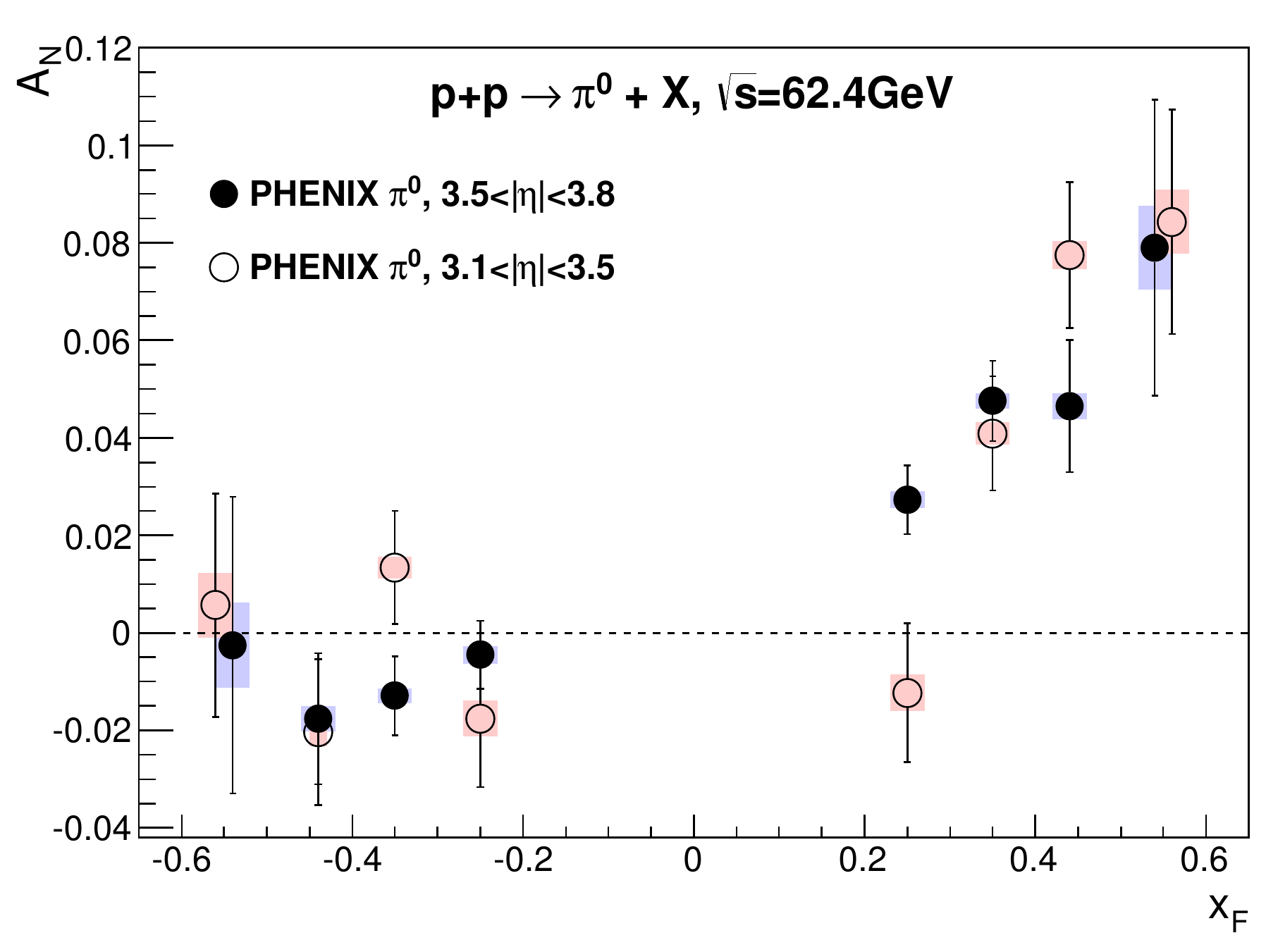}
\hskip 0.15in
\includegraphics[width=0.45\textwidth]{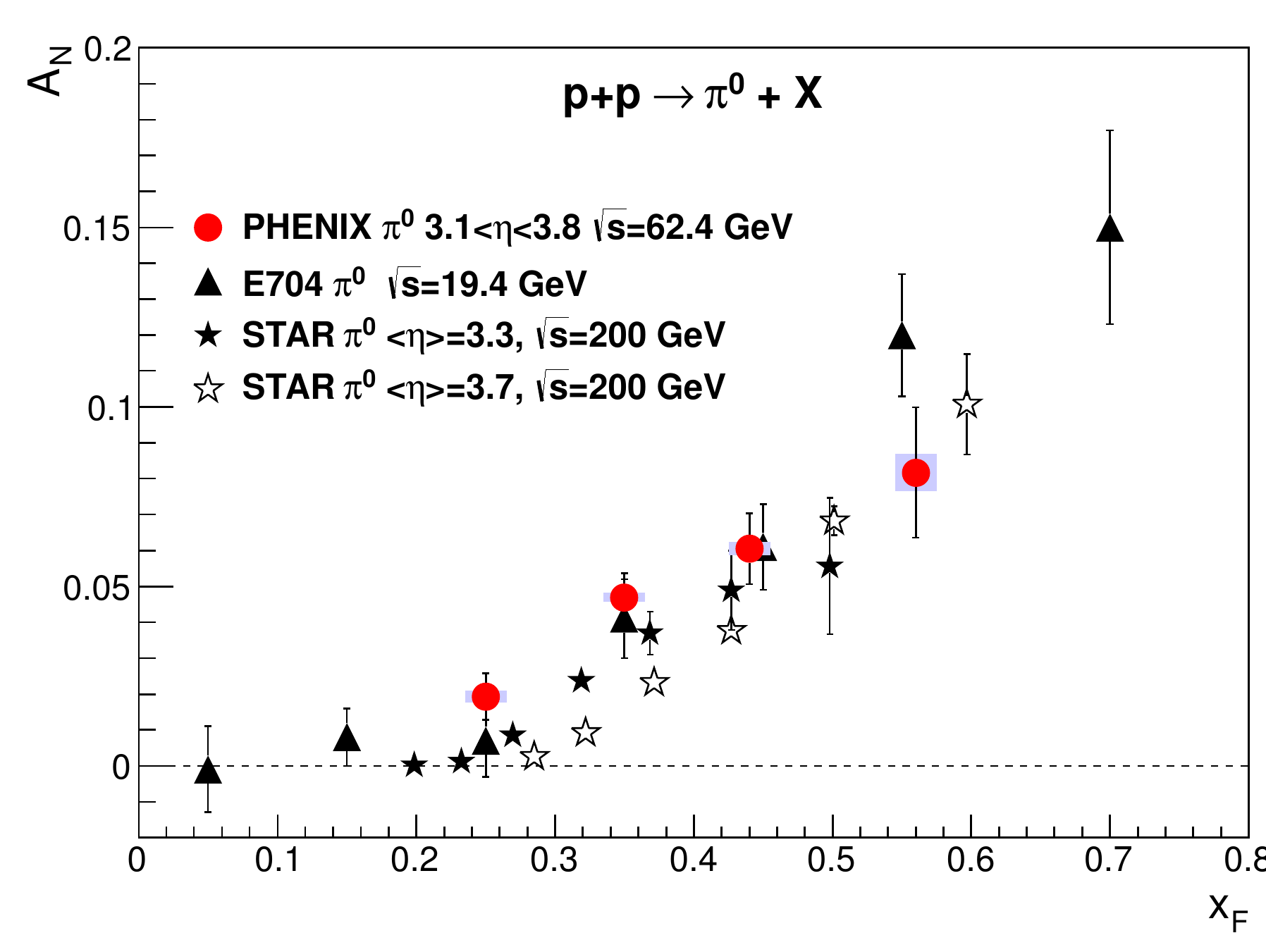}
\vspace{-0.1in}
\caption{$A_N^{\pi^0}$ as a function of $x_F$ at $\sqrt{s}=62.4$~GeV (Left) measured by PHENIX Experiment, and at various center-of-mass energies in hadronic collisions (Right).
\label{fig:an-pi0-rhic}}
\vspace{-0.1in}
\end{center}
\end{figure}

The left plot in Fig.~\ref{fig:an-pi0-rhic} shows $A_{N}^{\pi^0}$ at $\sqrt{s}=62.4$~GeV
measured by PHENIX Collaboration with the PHENIX detector at RHIC \cite{Adare:2013ekj}.  
While there is a significant, nonzero asymmetry rising with $x_F > 0$ 
in the forward direction of the polarized proton beam,
no such behavior can be seen at negative $x_F < 0$ where
the asymmetries are consistent with zero.  The typical transverse momentum 
of these data is relatively low, over the range of $0.5<p_T< 1$~GeV \cite{Adare:2013ekj}.  
In Fig.~\ref{fig:an-pi0-rhic} (Right), the PHENIX data of $A_{N}^{\pi^0}$ 
at $\sqrt{s}=62.4$~GeV are compared with the measurements by Fermilab E704 at 
$\sqrt{s}=19.4$~GeV and those by STAR Collaboration at $\sqrt{s}=200$~GeV.  
Although these measurements were carried out with slightly different detector 
acceptances, there is a general agreement between the $x_F$ dependence
of nonvanishing asymmetries.  The observed asymmetries appear to be independent 
of the center-of-mass energy of the collisions, which varies over one order of 
magnitude from $ \sqrt{s} =19.4$~\text{GeV} to $200$~GeV.  
The data set plotted in Fig.~\ref{fig:an-pi0-rhic} (Right) also covered an interesting  
range of transverse momentum of the observed pions from a semi-hard scale at
$p_T\sim 0.5$~GeV to a hard scale at $p_T> 2$~GeV for the STAR data 
\cite{Abelev:2008af}.
%%%%%%%%%%%%%%%%%%%%%%%%%%%%%%%%%%%%%%%%%%%%%%%%
\begin{figure}[!h]
\begin{center}
\vspace{-0.1in}
\includegraphics[width=0.45\textwidth,height=2.3in]{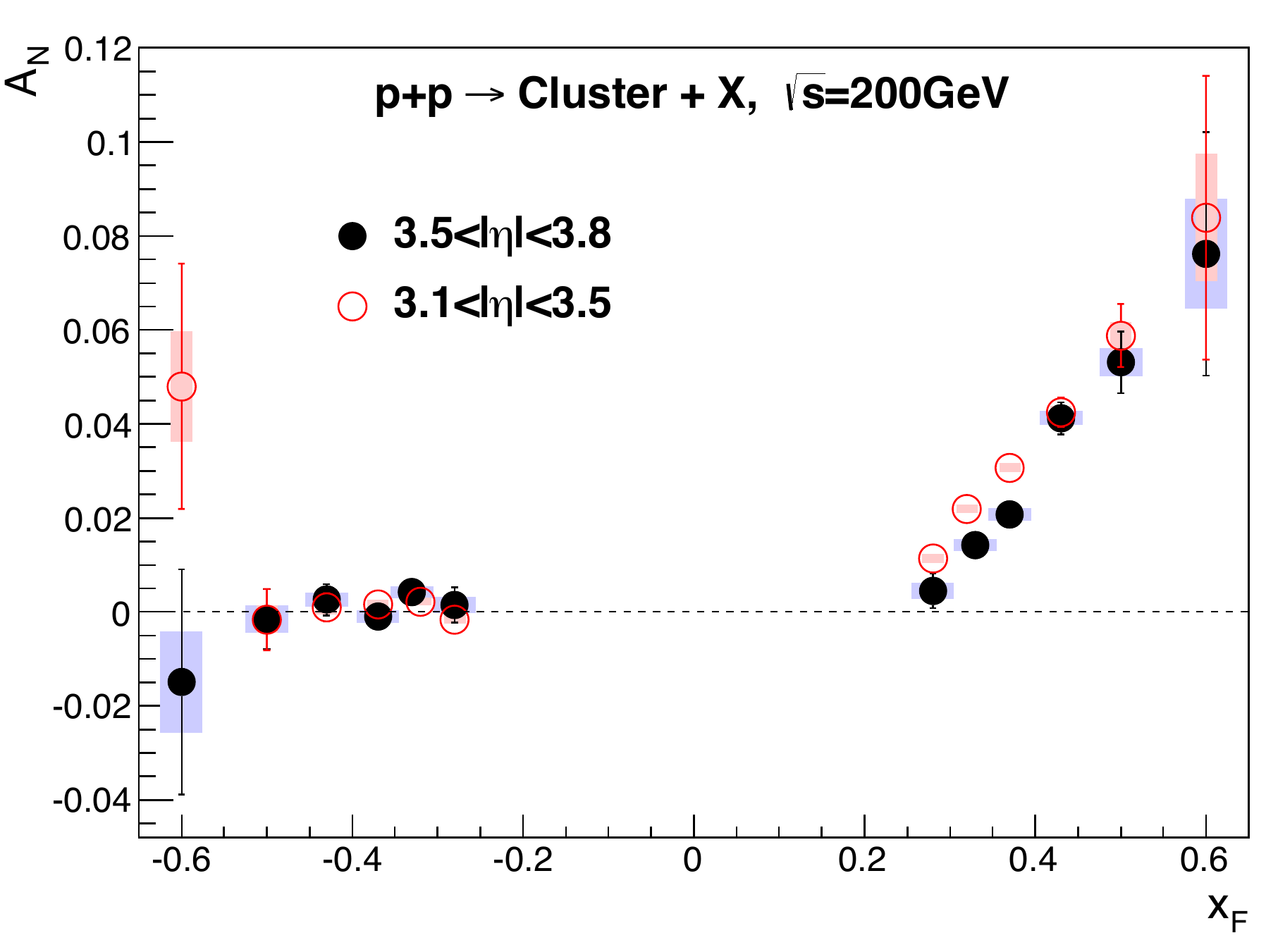}
\hskip 0.1in
\includegraphics[width=0.45\textwidth,height=2.3in]{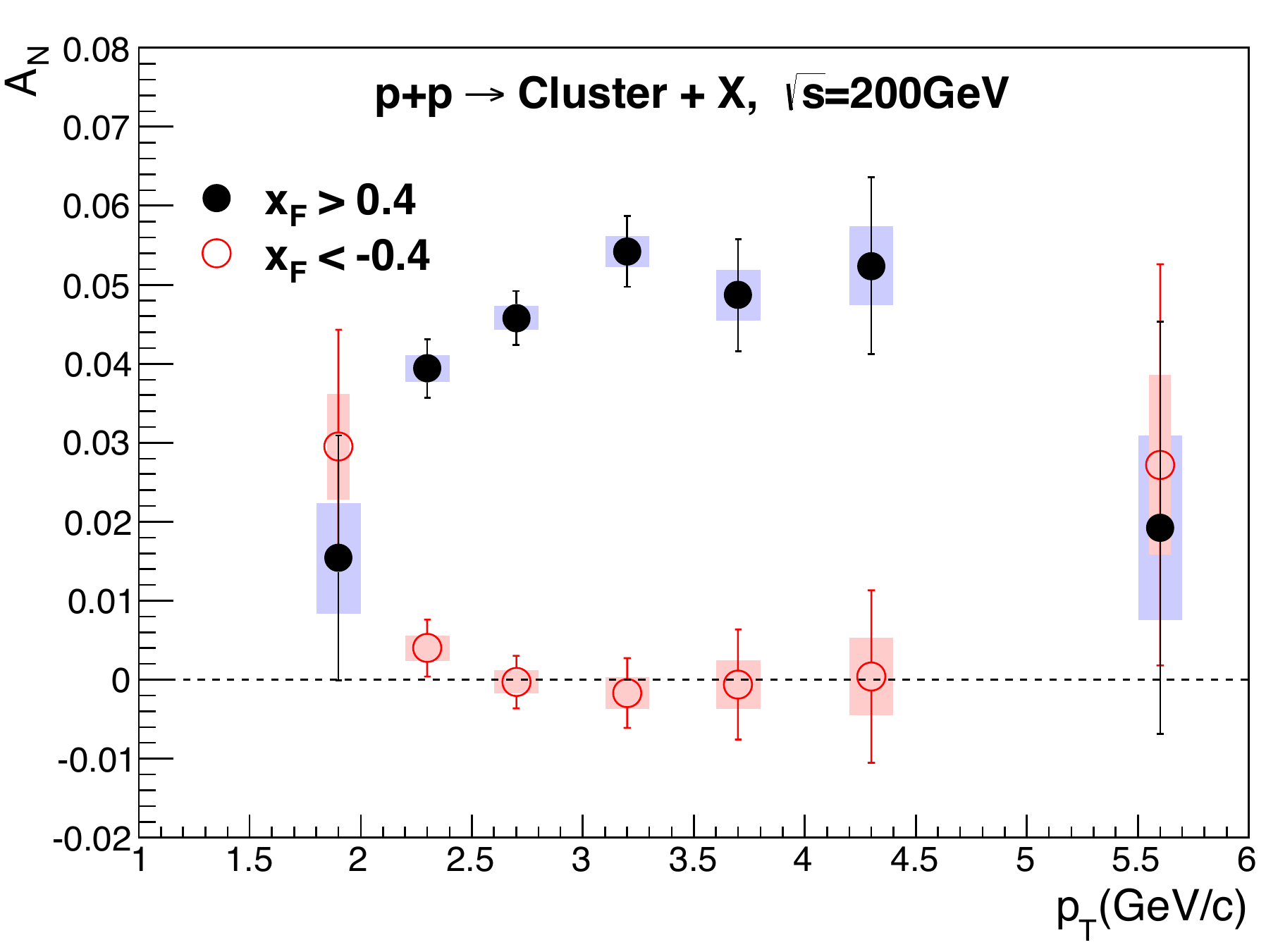}
\caption{$A_N^{\rm Cluster}$ as a function of the cluster's Feynman parameter $x_F$ (Left) and transverse momentum $p_T$ (Right)
measured by PHENIX Collaboration at $\sqrt{s}=200$~GeV at RHIC \cite{Adare:2013ekj}.  
\label{fig:an-cluster-rhic}}
\vspace{-0.2in}
\end{center}
\end{figure}

By resolving the two electromagnetic clusters from the two photons of 
$\pi^{0}\rightarrow\gamma+\gamma$ decay, PHENIX detector 
is capable of detecting $\pi^0$ with its energy, $E_{\pi^0}\lesssim$ 20~GeV.  
However, with increasing energy, the opening angle between the two photons 
becomes so small that their electromagnetic clusters fully merge in the PHENIX 
detector. The reconstruction of $\pi^0$ from the two-gamma decay mode
limits the $x_{F}$ range to below 0.2 at $\sqrt{s}=200$~GeV, where 
transverse SSAs are small.  To overcome this limitation,  
the data analysis of transverse SSAs is also done for inclusive 
electromagnetic clusters \cite{Adare:2013ekj}.  The clusters are dominated 
from $\pi^0\to \gamma+\gamma$, with some contributions from direct and other 
photons and decay of $\eta$ and other charged particles.  The detailed discussion
on the decomposition of the clusters can be found in Ref.~\cite{Adare:2013ekj}.

The left plot in Fig.~\ref{fig:an-cluster-rhic} summarizes the $x_{F}$-dependence 
of the cluster $A_{N}$ for two different pseudorapidity ranges.  
Within statistical uncertainties the asymmetries in the forward 
direction of the polarized proton $A_{N}$ rises almost linearly with $x_{F}$,
while the asymmetries in the backward direction $x_{F}<0$ are found to 
be consistent with zero.  The non-vanish asymmetries in the forward region are 
of similar size compared to the $\pi^0$ asymmetries at other center-of-mass 
energies as shown in Fig.~\ref{fig:an-pi0-rhic} (Right), which should not be too surprising
since the inclusive electromagnetic clusters in this momentum regime are mainly 
formed by the photons from $\pi^0$ decay.  

The right plot in Fig.~\ref{fig:an-cluster-rhic} presents the cluster $A_{N}$, 
as a function of transverse momentum $p_{T}$ for values of $|x_{F}|>0.4$ \cite{Adare:2013ekj}. 
The asymmetry rises smoothly and then seems to saturate above 
$p_{T}>3$~GeV/$c$.  When $p_T=0$, $A_N$ should vanish from the symmetry,
and is expected to rise with $p_T$ when $p_T$ is relatively small.  Since transverse
SSAs of the single inclusive hadron production are intrinsically a power suppressed 
observable at large $p_T$, the $A_N$ is expected to decrease 
when $p_T$ becomes sufficiently large \cite{Qiu:1991pp,Ji:2006ub}.
Explicit calculations of $A_N$ in collinear factorization approach indicates,
$A_N \propto p_T/\hat{u}$, and consequently, the asymmetry does not fall
as quickly as $1/p_T$ in the forward region 
when $x_F \gtrsim x_T = 2p_T/\sqrt{s}$ \cite{Qiu:1998ia}. 
The $p_T$ dependence of the $A_N$ should be sensitive to the underline 
dynamics generating the transverse SSAs.  
Again, asymmetries in the negative $x_{F}$ region are found to 
be consistent with zero within statistical uncertainties.
The transverse SSAs of single inclusive hadron or inclusive cluster measured
by PHENIX Experiment are consistent with those measured by STAR collobration
at RHIC.  

%======================================================================
\subsection{$A_N$ of Drell-Yan ($\gamma^*$ and $W/Z$) production}
\label{subsec:w}
%======================================================================

The observed transverse SSAs of single inclusive hadron production could be generated 
by the spin-motion correlations and/or coherent multiple interactions before and/or 
after the hard collision \cite{Qiu:1998ia}.  That is, in the terminology of 
TMD factorization approach, both the Sivers-like (before) and the Collins-like (after) 
effects are responsible for the observed asymmetries.  Assuming the correlation before
the hard collision dominates the observed asymmetry in the inclusive pion production 
could actually lead to a sign puzzle for the relation between the Sivers function and 
the corresponding twist-3 quark-gluon correlation functions \cite{Kang:2011hk}.  
Apparently, the Collins-like correlations after the hard collision is very important for
understanding the transverse SSAs in $p$-$p$ collisions, as well as resolving 
the sign puzzle \cite{Kanazawa:2014dca}. Without being able to 
disentangle different effects, it is difficult, if not impossible, to fully understand the 
physics and the mechanism to generate the novel phenomena of transverse SSAs.

Unlike the single inclusive hadron production in hadronic collision, Drell-Yan massive 
lepton pair production, either via a virtual photon or a $W/Z$ boson, could have 
two natural and distinctive momentum scales:  the invariant mass of the lepton pair, 
$Q$ ($\sim M_{W/Z}$) and the transverse momentum of the pair, $q_T$.  
The most events have the strong ordering of these two scales, $Q\gg q_T$, 
which is natrual for applying the TMD factorization approach to probe the 
transverse motion of quarks directly.  Since the vector bosons
or their decay leptons do not interact strongly after they are produced, Drell-Yan
process is ideal for study the initial-state TMDs, such as Sivers functions \cite{Peng:2014hta}.

One of the key differences between the QCD collinear factorization approach and 
QCD TMD factorization approach to the high energy scattering process 
is the universality of the non-perturbative long-distance physics.  
Within the collinear factorization approach, all non-perturbative long-distance
physics are represented by universal (process independent) hadronic matrix elements 
whose operators are made of quark/gluon correlators that are effectively {\it localized} in space 
to the size of the hard collision $\sim 1/Q\ll 1/\text{fm}$.
The non-perturbative long-distance physics in the TMD factorization approach is represented 
by TMDs.  The existing definitions of TMDs involve quark-gluon correlators with 
{\it non-local} gauge links covering infinite size in space \cite{Collins:2012uy}.  
It is the non-local nature of the TMDs that led to their potential process-dependence,
such as the sign change of the Sivers function extracted from SIDIS in comparison with
that extracted from the Drell-Yan process \cite{Collins:2002kn}.  
This non-universality is a fundamental prediction from the gauge invariance of QCD
and the TMD factorization approach. 
The experimental test of this sign change is one of the open questions in hadronic
physics and will provide a direct verification of QCD TMD factorization.
%%%%%%%%%%%%%%%%%%%%%%%%%%%%%%%%%%%%%%%%%%%%%%%%
\begin{figure}[!h]
\begin{center}
\includegraphics[width=0.98\textwidth]{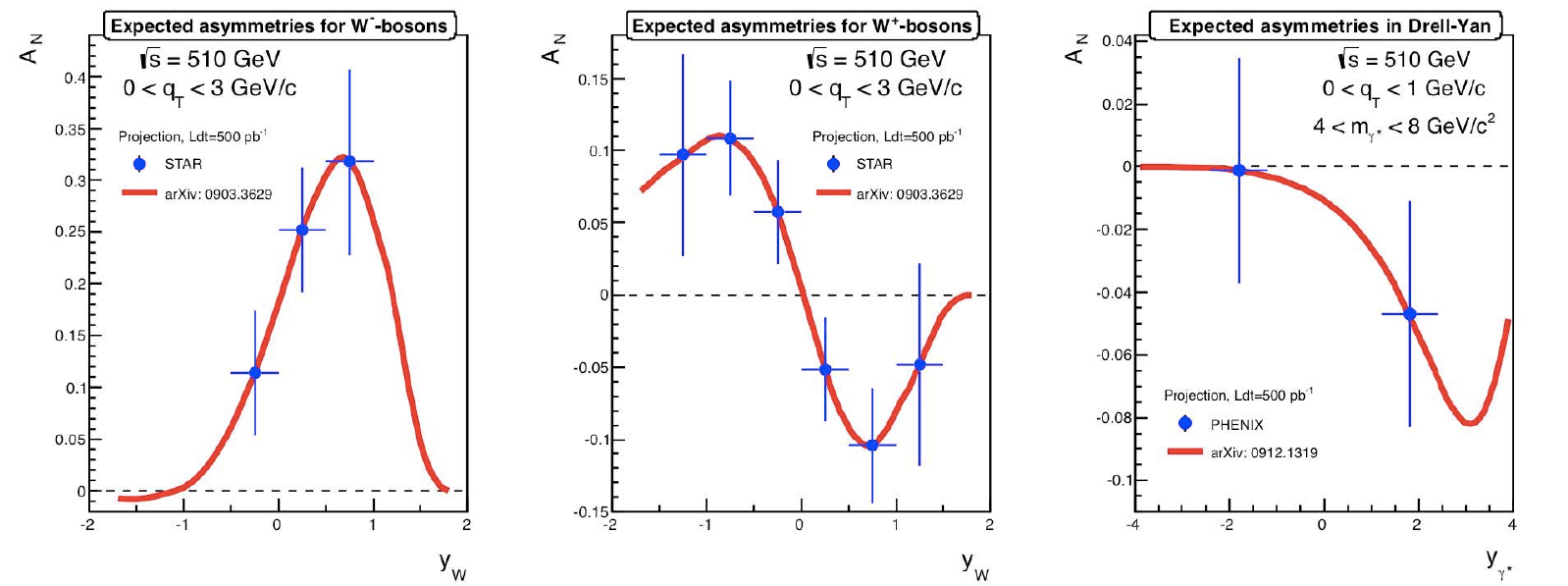}
\caption{$A_N$ of Drell-Yan like process with a reconstructed $W^+$ (Left), $W^-$ (Middle), or a virtual photon (Right), as a function of vector boson's rapidity in $\vec{p}$-$p$ collisions at RHIC.   
Also plotted are theoretical predictions based on Sivers functions extracted from low
energy SIDIS measurements \cite{Kang:2009sm}.
\label{fig:an-w-rhic}}
\end{center}
\end{figure}

With the high energy polarized proton beams, RHIC can produce $W$ and $Z^0$ 
with relatively low transverse momentum $q_T\ll M_{W/Z}$. 
The transverse spin program at RHIC has the unique advantage to measure the Sivers
effect in the Drell-Yan like process at a true hard scale -- mass of $W/Z$ bosons.
Figure~\ref{fig:an-w-rhic} shows the expected uncertainties for transverse SSAs 
of reconstructed $W^\pm$ (Left, Middle) and Drell-Yan production (Right) 
from STAR and PHENIX measurements \cite{Aschenauer:2013woa}.
Also plotted curves are theoretical predictions based on our knowledges of 
Sivers functions from SIDIS measurements \cite{Kang:2009sm}.  These asymmetries
provide an essential test for the fundamental QCD prediction of 
a sign change of the Sivers function in hadronic collisions with
respect to that in SIDIS, as well as a test of our understanding of 
scale dependence (or evolution) of the Sivers functions.

Transverse SSAs and their $x_F$ and $p_T$ dependence in hadronic collisions 
are well-established, and they are emergent phenomena QCD dynamics.  
Along with the transverse spin programs at JLab12 and at COMPASS,
the transverse spin program at RHIC with its PHENIX and STAR detectors 
opens up a new domain of QCD dynamics, sensitive to the confined motion, 
as well as the quantum correlation between the motion and intrinsic particle property, 
such as spin, of quarks and gluons inside a QCD bound state, like the proton.  
It also provides controllable and promising access to the transverse structure of the proton.

%%%%%%%%%%%%%%%%%%%%%%%%%%%%%%%%%%%%%%%%%%%%%%%%
\section{Summary and outlook}
\label{sec:summary}
%%%%%%%%%%%%%%%%%%%%%%%%%%%%%%%%%%%%%%%%%%%%%%%%

Proton is a dynamic system of confined quarks and gluons.   
Proton itself is an emergent phenomenon of QCD dynamics.    
Proton structure is dynamic as well.  Probability distributions 
to find quarks and gluons, and their correlations inside the proton 
are fundamental, corresponding to various aspects of the 
dynamic and complex system of the proton.  Our knowledge of 
proton's internal structure is still very limited, although we 
have learned a lot with the advances of accelerator and detector technologies,
as well as theoretical breakthroughs to identify the controllable 
probes to ``connect'' the quarks and gluons inside the proton
to the leptons and hadrons observed in modern detectors.

RHIC, with the only polarized proton-proton collider in the world
and the PHENIX and STAR detectors, has helped address 
open questions concerning the proton structure, and identified 
new open questions, puzzles and challenges.  
The PHENIX Experiment, along with STAR Experiment, 
has played critical roles in defining the RHIC physics program 
and in extracting valuable information 
for determining proton's internal structure.  
The PHENIX Experiment helps determine the helicity structure of the proton,
in particular, the extraction of gluon and sea quark helicity structure
inside a polarized fast moving proton, by measuring $A_{LL}$ of single 
inclusive hadron production, as well as the parity violating $A_L$ of 
$W^\pm$ production.  By measuring $A_N$ of many observables, including the production of 
$\pi^0$, $\eta$, direct photon, $W/Z$ bosons, $J/\psi$, and the innovative inclusive 
electromagnetic clusters, the PHENIX Experiment help define and explore 
the richness and excitements of the RHIC transverse spin program.

With the upcoming $p$-$A$ run at RHIC, the PHENIX Experiment will
help open a new frontier of QCD to explore the fundamental nuclear structure 
in terms of quarks and gluons.  With the help of the polarized proton beam, 
the PHENIX Experiment could search for the imprint of spin correlations
among the nucleons inside a nucleus, without seeing nucleon, but, 
quarks and gluons.  With the future upgrades of the PHENIX detector
to sPHENIX and to fsPHENIX, the PHENIX Experiment could be even
more powerful in delivering new and accurate measurements helping
explore the internal structure of the proton.

%%%%%%%%%%%%%%%%%%%%%%%%%%%%%%%%%%%%%%%%%%%%%%%%
\section*{Acknowledgment}

We thank Abhay Deshpande for helpful discussions concerning PHENIX measurements. 
This work was supported in part by the U. S. Department of Energy under Contract 
No.~DE-AC02-98CH10886, and the National Science Foundation under Grants 
No.~PHY-0969739 and No. PHY-1316617.

%%%%%%%%%%%%%%%%%%%%%%%%%%%%%%%%%%%%%%%%%%%%%%%%
\bibliographystyle{ptephy}
\bibliography{eic_white_paper}

\begin{thebibliography}{10}

\bibitem{Brambilla:2014jmp}
N.~Brambilla, S.~Eidelman, P.~Foka, S.~Gardner, A.S. Kronfeld, et~al.,
  Eur.Phys.J., {\bf C74}(10), 2981 (2014),  {{arXiv:1404.3723}}.

\bibitem{Bloom:1969kc}
Elliott~D. Bloom, D.H. Coward, H.C. DeStaebler, J.~Drees, Guthrie Miller,
  et~al., Phys.Rev.Lett., {\bf 23}, 930--934 (1969).

\bibitem{Breidenbach:1969kd}
Martin Breidenbach, Jerome~I. Friedman, Henry~W. Kendall, Elliott~D. Bloom,
  D.H. Coward, et~al., Phys.Rev.Lett., {\bf 23}, 935--939 (1969).

\bibitem{Fritzsch:1973pi}
H.~Fritzsch, Murray Gell-Mann, and H.~Leutwyler, Phys.Lett., {\bf B47},
  365--368 (1973).

\bibitem{RHIChomepage}
{RHIC homepage:\ http://www.bnl.gov/rhic/} ().

\bibitem{Gyulassy:2004zy}
M.~Gyulassy and L.~McLerran, Nucl. Phys., {\bf A750}, 30 (2005).

\bibitem{Adamczyk:2014ozi}
L.~Adamczyk et~al. (2014),  {{arXiv:1405.5134}}.

\bibitem{Adare:2014hsq}
A.~Adare et~al., Phys.Rev., {\bf D90}, 012007 (2014),  {{arXiv:1402.6296}}.

\bibitem{Adamczyk:2014xyw}
L.~Adamczyk et~al., Phys.Rev.Lett., {\bf 113}, 072301 (2014),
  {{arXiv:1404.6880}}.

\bibitem{Adare:2010xa}
A.~Adare et~al., Phys. Rev. Lett., {\bf 106}, 062001 (2011),
  {{arXiv:1009.0505}}.

\bibitem{Abelev:2008af}
B.I. Abelev et~al., Phys.Rev.Lett., {\bf 101}, 222001 (2008),
  {{arXiv:0801.2990}}.

\bibitem{Adare:2013ekj}
A.~Adare et~al., Phys.Rev., {\bf D90}, 012006 (2014),  {{arXiv:1312.1995}}.

\bibitem{Gao:2013xoa}
Jun Gao, Marco Guzzi, Joey Huston, Hung-Liang Lai, Zhao Li, et~al., Phys.Rev.,
  {\bf D89}(3), 033009 (2014),  {{arXiv:1302.6246}}.

\bibitem{Martin:2009iq}
A.D. Martin, W.J. Stirling, R.S. Thorne, and G.~Watt, Eur.Phys.J., {\bf C63},
  189--285 (2009),  {{arXiv:0901.0002}}.

\bibitem{Ball:2011mu}
Richard~D. Ball, Valerio Bertone, Francesco Cerutti, Luigi Del~Debbio, Stefano
  Forte, et~al., Nucl.Phys., {\bf B849}, 296--363 (2011),  {{arXiv:1101.1300}}.

\bibitem{Alekhin:2013nda}
S.~Alekhin, J.~Bluemlein, and S.~Moch, Phys.Rev., {\bf D89}, 054028 (2014),
  {{arXiv:1310.3059}}.

\bibitem{Ashman:1987hv}
J.~Ashman et~al., Phys. Lett., {\bf B206}, 364 (1988).

\bibitem{Ashman:1989ig}
J.~Ashman et~al., Nucl. Phys., {\bf B328}, 1 (1989).

\bibitem{Aschenauer:2013woa}
E.C. Aschenauer, A.~Bazilevsky, K.~Boyle, K.O. Eyser, R.~Fatemi, et~al. (2013),
   {{arXiv:1304.0079}}.

\bibitem{Jaffe:1989jz}
R.~L. Jaffe and Aneesh Manohar, Nucl. Phys., {\bf B337}, 509--546 (1990).

\bibitem{Ji:1996ek}
Xiang-Dong Ji, Phys. Rev. Lett., {\bf 78}, 610--613 (1997),
  {{hep-ph/9603249}}.

\bibitem{Wakamatsu:2014zza}
Masashi Wakamatsu, Int.J.Mod.Phys., {\bf A29}, 1430012 (2014),
  {{arXiv:1402.4193}}.

\bibitem{Ji:2012sj}
Xiangdong Ji, Xiaonu Xiong, and Feng Yuan, Phys. Rev. Lett., {\bf 109}, 152005
  (2012),  {{arXiv:1202.2843}}.

\bibitem{CSS-FAC}
John~C. Collins, Davison~E. Soper, and George~F. Sterman, Adv.Ser.Direct.High
  Energy Phys., {\bf 5}, 1--91 (1988),  {{arXiv:hep-ph/0409313}}.

\bibitem{Dokshitzer:1977sg}
Yuri~L. Dokshitzer, Sov. Phys. JETP, {\bf 46}, 641--653 (1977).

\bibitem{Gribov:1972ri}
V.N. Gribov and L.N. Lipatov, Sov. J. Nucl. Phys., {\bf 15}, 438--450 (1972).

\bibitem{Altarelli:1977zs}
Guido Altarelli and G.~Parisi, Nucl. Phys., {\bf B126}, 298 (1977).

\bibitem{Zijlstra:1993sh}
E.B. Zijlstra and W.L. van Neerven, Nucl. Phys., {\bf B417}, 61--100 (1994).

\bibitem{Mertig:1995ny}
R.~Mertig and W.L. van Neerven, Z. Phys., {\bf C70}, 637--654 (1996),
  {{arXiv:hep-ph/9506451}}.

\bibitem{Vogelsang:1995vh}
Werner Vogelsang, Phys. Rev., {\bf D54}, 2023--2029 (1996),
  {{arXiv:hep-ph/9512218}}.

\bibitem{Vogt:2008yw}
A.~Vogt, S.~Moch, M.~Rogal, and J.A.M. Vermaseren, Nucl. Phys. Proc. Suppl.,
  {\bf 183}, 155--161 (2008),  {{arXiv:0807.1238}}.

\bibitem{Accardi:2012qut}
A.~Accardi, J.L. Albacete, M.~Anselmino, N.~Armesto, E.C. Aschenauer, et~al.
  (2012),  {{arXiv:1212.1701}}.

\bibitem{Adler:2004ps}
S.S. Adler et~al., Phys. Rev. Lett., {\bf 93}, 202002 (2004),
  {{arXiv:hep-ex/0404027}}.

\bibitem{Adare:2007dg}
A.~Adare et~al., Phys. Rev., {\bf D76}, 051106 (2007),  {{arXiv:0704.3599}}.

\bibitem{Adare:2008px}
A.~Adare et~al., Phys. Rev. Lett., {\bf 103}, 012003 (2009),
  {{arXiv:0810.0694}}.

\bibitem{Adare:2008qb}
A.~Adare et~al., Phys.Rev., {\bf D79}, 012003 (2009),  {{arXiv:0810.0701}}.

\bibitem{Adare:2010cy}
A.~Adare et~al., Phys.Rev., {\bf D83}, 032001 (2011),  {{arXiv:1009.6224}}.

\bibitem{Abelev:2006uq}
B.I. Abelev et~al., Phys. Rev. Lett., {\bf 97}, 252001 (2006),
  {{arXiv:hep-ex/0608030}}.

\bibitem{Abelev:2007vt}
B.I. Abelev et~al., Phys. Rev. Lett., {\bf 100}, 232003 (2008),
  {{arXiv:0710.2048}}.

\bibitem{Sarsour:2009zd}
M.~Sarsour, AIP Conf. Proc., {\bf 1149}, 389--392 (2009),  {{arXiv:0901.4061}}.

\bibitem{Djawotho:2011zz}
Pibero Djawotho, J. Phys. Conf. Ser., {\bf 295}, 012061 (2011).

\bibitem{Abelev:2009pb}
B.I. Abelev et~al., Phys. Rev., {\bf D80}, 111108 (2009),  {{arXiv:0911.2773}}.

\bibitem{deFlorian:2008mr}
Daniel de~Florian, Rodolfo Sassot, Marco Stratmann, and Werner Vogelsang, Phys.
  Rev. Lett., {\bf 101}, 072001 (2008),  {{arXiv:0804.0422}}.

\bibitem{deFlorian:2009vb}
Daniel de~Florian, Rodolfo Sassot, Marco Stratmann, and Werner Vogelsang, Phys.
  Rev., {\bf D80}, 034030 (2009),  {{arXiv:0904.3821}}.

\bibitem{Blumlein:2010rn}
Johannes Blumlein and Helmut Bottcher, Nucl.Phys., {\bf B841}, 205--230 (2010),
   {{arXiv:1005.3113}}.

\bibitem{Ball:2013lla}
Richard~D. Ball et~al., Nucl.Phys., {\bf B874}, 36--84 (2013),
  {{arXiv:1303.7236}}.

\bibitem{Ball:2012cx}
Richard~D. Ball, Valerio Bertone, Stefano Carrazza, Christopher~S. Deans, Luigi
  Del~Debbio, et~al., Nucl.Phys., {\bf B867}, 244--289 (2013),
  {{arXiv:1207.1303}}.

\bibitem{Leader:2010rb}
Elliot Leader, Aleksander~V. Sidorov, and Dimiter~B. Stamenov, Phys.Rev., {\bf
  D82}, 114018 (2010),  {{arXiv:1010.0574}}.

\bibitem{deFlorian:2014yva}
Daniel de~Florian, Rodolfo Sassot, Marco Stratmann, and Werner Vogelsang,
  Phys.Rev.Lett., {\bf 113}, 012001 (2014),  {{arXiv:1404.4293}}.

\bibitem{Peng:2014hta}
Jen-Chieh Peng and Jian-Wei Qiu, Prog.Part.Nucl.Phys., {\bf 76}, 43--75 (2014),
   {{arXiv:1401.0934}}.

\bibitem{Hawker:1998ty}
E.A. Hawker et~al., Phys.Rev.Lett., {\bf 80}, 3715--3718 (1998),
  {{arXiv:hep-ex/9803011}}.

\bibitem{Peng:1998pa}
J.C. Peng et~al., Phys.Rev., {\bf D58}, 092004 (1998),
  {{arXiv:hep-ph/9804288}}.

\bibitem{Towell:2001nh}
R.S. Towell et~al., Phys.Rev., {\bf D64}, 052002 (2001),
  {{arXiv:hep-ex/0103030}}.

\bibitem{Corliss:2014dha}
Ross Corliss, Phys.Part.Nucl., {\bf 45}, 70--72 (2014).

\bibitem{deFlorian:2010aa}
Daniel de~Florian and Werner Vogelsang, Phys. Rev., {\bf D81}, 094020 (2010),
  {{arXiv:1003.4533}}.

\bibitem{Gal:2014fha}
C.~Gal, Phys.Part.Nucl., {\bf 45}, 76--78 (2014).

\bibitem{Belitsky:2003nz}
Andrei~V. Belitsky, Xiang-Dong Ji, and Feng Yuan, Phys. Rev., {\bf D69}, 074014
  (2004),  {{arXiv:hep-ph/0307383}}.

\bibitem{Hillery:1983ms}
M.~Hillery, R.F. O'Connell, M.O. Scully, and Eugene~P. Wigner, Phys. Rept.,
  {\bf 106}, 121--167 (1984).

\bibitem{Mueller:2014hsa}
Dieter Mueller (2014),  {{arXiv:1405.2817}}.

\bibitem{Burkardt:2005km}
Matthias Burkardt and Gunar Schnell, Phys. Rev., {\bf D74}, 013002 (2006),
  {{hep-ph/0510249}}.

\bibitem{Sivers:1989cc}
Dennis~W. Sivers, Phys.Rev., {\bf D41}, 83 (1990).

\bibitem{Collins:1992kk}
John~C. Collins, Nucl.Phys., {\bf B396}, 161--182 (1993),
  {{arXiv:hep-ph/9208213}}.

\bibitem{Kane:1978nd}
Gordon~L. Kane, J.~Pumplin, and W.~Repko, Phys.Rev.Lett., {\bf 41}, 1689
  (1978).

\bibitem{Qiu:1990cu}
Jian-Wei Qiu and George~F. Sterman, AIP Conf.Proc., {\bf 223}, 249--254 (1991).

\bibitem{Brodsky:2002cx}
Stanley~J. Brodsky, Dae~Sung Hwang, and Ivan Schmidt, Phys. Lett., {\bf B530},
  99--107 (2002).

\bibitem{Collins:2002kn}
John~C. Collins, Phys. Lett., {\bf B536}, 43--48 (2002).

\bibitem{Efremov:1981sh}
A.V. Efremov and O.V. Teryaev, Sov.J.Nucl.Phys., {\bf 36}, 140 (1982).

\bibitem{Efremov:1984ip}
A.V. Efremov and O.V. Teryaev, Phys.Lett., {\bf B150}, 383 (1985).

\bibitem{Qiu:1991pp}
Jian-Wei Qiu and George~F. Sterman, Phys.Rev.Lett., {\bf 67}, 2264--2267
  (1991).

\bibitem{Qiu:1991wg}
Jian-Wei Qiu and George~F. Sterman, Nucl.Phys., {\bf B378}, 52--78 (1992).

\bibitem{Qiu:1998ia}
Jian-Wei Qiu and George~F. Sterman, Phys.Rev., {\bf D59}, 014004 (1998),
  {{arXiv:hep-ph/9806356}}.

\bibitem{Kouvaris:2006zy}
Chris Kouvaris, Jian-Wei Qiu, Werner Vogelsang, and Feng Yuan, Phys.Rev., {\bf
  D74}, 114013 (2006),  {{arXiv:hep-ph/0609238}}.

\bibitem{Qiu:1993af}
Jian-Wei Qiu and George~F. Sterman, In *Brookhaven 1993, Future directions in
  particle and nuclear physics at multi-GeV hadron beam facilities*, Brookhaven
  National Lab, Upton, NY (1993).

\bibitem{Ji:2006ub}
Xiangdong Ji, Jian-Wei Qiu, Werner Vogelsang, and Feng Yuan, Phys. Rev. Lett.,
  {\bf 97}, 082002 (2006).

\bibitem{Ji:2006vf}
Xiangdong Ji, Jian-Wei Qiu, Werner Vogelsang, and Feng Yuan, Phys.Rev., {\bf
  D73}, 094017 (2006),  {{arXiv:hep-ph/0604023}}.

\bibitem{Ji:2006br}
Xiangdong Ji, Jian-Wei Qiu, Werner Vogelsang, and Feng Yuan, Phys.Lett., {\bf
  B638}, 178--186 (2006),  {{arXiv:hep-ph/0604128}}.

\bibitem{Koike:2007dg}
Yuji Koike, Werner Vogelsang, and Feng Yuan, Phys.Lett., {\bf B659}, 878--884
  (2008),  {{arXiv:0711.0636}}.

\bibitem{Bacchetta:2008xw}
Alessandro Bacchetta, Daniel Boer, Markus Diehl, and Piet~J. Mulders, JHEP,
  {\bf 08}, 023 (2008).

\bibitem{Kang:2011hk}
Zhong-Bo Kang, Jian-Wei Qiu, Werner Vogelsang, and Feng Yuan, Phys.Rev., {\bf
  D83}, 094001 (2011),  {{arXiv:1103.1591}}.

\bibitem{Kanazawa:2014dca}
Koichi Kanazawa, Yuji Koike, Andreas Metz, and Daniel Pitonyak, Phys.Rev., {\bf
  D89}(11), 111501 (2014),  {{arXiv:1404.1033}}.

\bibitem{Collins:2012uy}
John~C. Collins and Ted~C. Rogers, Phys.Rev., {\bf D87}(3), 034018 (2013),
  {{arXiv:1210.2100}}.

\bibitem{Kang:2009sm}
Zhong-Bo Kang and Jian-Wei Qiu, Phys.Rev., {\bf D81}, 054020 (2010),
  {{arXiv:0912.1319}}.

\end{thebibliography}

%%%%%%%%%%%%%%%%%%%%%%%%%%%%%%%%%%%%%%%%%%%%%%%%
\end{document}